\documentclass[12pt]{article}

\renewcommand{\thefootnote}{\fnsymbol{footnote}}

\usepackage{amsbsy,amssymb,latexsym,amsfonts,amsmath}
\usepackage{mathrsfs}
\usepackage{bm}
\usepackage{here}
\usepackage[dvipdfmx]{graphicx}
\usepackage{color}

\begin{document}
%
%
\begin{titlepage}
\begin{flushright}
\normalsize
~~~~
NITEP 21\\
OCU-PHYS 503\\
May, 2019 \\
\end{flushright}

\vspace{15pt}

\begin{center}
{\LARGE  Exponentially suppressed cosmological constant with enhanced gauge symmetry in heterotic interpolating models} \\
\end{center}

\vspace{23pt}

\begin{center}
{ H. Itoyama$^{a, b,c}$\footnote{e-mail: itoyama@sci.osaka-cu.ac.jp},
  Sota Nakajima$^b$\footnote{e-mail: sotanaka@sci.osaka-cu.ac.jp}   }\\

%
\vspace{10pt}
%

$^a$\it Nambu Yoichiro Institute of Theoretical and Experimental Physics (NITEP),\\
Osaka City University\\
\vspace{5pt}

$^b$\it Department of Mathematics and Physics, Graduate School of Science,\\
Osaka City University\\
\vspace{5pt}

$^c$\it Osaka City University Advanced Mathematical Institute (OCAMI)

\vspace{5pt}

3-3-138, Sugimoto, Sumiyoshi-ku, Osaka, 558-8585, Japan \\

\end{center}
%
\vspace{15pt}
\begin{center}
Abstract\\
\end{center}
 A few nine-dimensional interpolating models with two parameters are constructed and the massless spectra are studied by considering compactification of heterotic strings on a twisted circle with Wilson line. It is found that there are some conditions between radius $R$ and Wilson line $A$ under which the gauge symmetry is enhanced. In particular, when the gauge symmetry is enhanced to $SO(18)\times SO(14)$, the cosmological constant is exponentially suppressed. We also construct a non-supersymmetric string model which is tachyon-free in all regions of moduli space and whose gauge symmetry involves $E_8$. 

\vfill

\end{titlepage}

\renewcommand{\thefootnote}{\arabic{footnote}}
\setcounter{footnote}{0}


\section{Introduction}

LHC experiments suggest that supersymmetry (SUSY) does not exist at low energy scale. It is, therefore, natural to consider the possibility that SUSY is broken at the string/Planck scale. For this reason, non-supersymmetric string models \cite{String Theories in Ten-Dimensions Without Space-Time Supersymmetry,An O(16) x O(16) Heterotic String,Nonsupersymmetric orbifolds}, in particular, the $SO(16)\times SO(16)$ heterotic string model which is the unique tachyon-free ten-dimensional non-supersymmetric model, are receiving more and more attention.  Non-supersymmetric string models, however, always have a problem of stability. Unlike the supersymmetric ones, the cosmological constant is non-vanishing. There are non-vanishing dilaton tadpoles which lead to vacuum instability. Thus, the desired model must both be non-supersymmetric and carry a very small cosmological constant. While several methods to construct such models have been proposed\cite{ATKIN-LEHNER SYMMETRY,The Failure of Atkin-lehner Symmetry for Lattice Compactified Strings,GENERALIZED ATKIN-LEHNER SYMMETRY,Model Building on Asymmetric Z(3) Orbifolds: Nonsupersymmetric Models,Non-supersymmetric Asymmetric Orbifolds with Vanishing Cosmological Constant,More on Non-supersymmetric Asymmetric Orbifolds with Vanishing Cosmological Constant,Vacuum energy cancellation in a nonsupersymmetric string,On vanishing two loop cosmological constants in nonsupersymmetric strings}, in this paper, we try to construct non-supersymmetric heterotic models with a small cosmological constant by focusing on so-called interpolating models\cite{Supersymmetry Restoration in the Compactified O(16) x O(16)-prime Heterotic String Theory,Strong / weak coupling duality relations for nonsupersymmetric string theories,Towards a nonsupersymmetric string phenomenology,Interpolations from supersymmetric to nonsupersymmetric strings and their properties}.

An interpolating model is a $(D-d)$-dimensional model that continuously relates two $D$-dimensional models. In this work, we restrict our attention to the case with $D = 10$ and $d = 1$ for simplicity. The method of constructing such models is as follows; We start from a ten-dimensional closed string model (called model $M_1$) and compactify this on a circle with a $\boldsymbol{Z}_2$ twist, which is nothing but the Scherk-Schwarz compactification \cite{Spontaneous Breaking of Supersymmetry Through Dimensional Reduction,Spontaneous Supersymmetry Breaking in Supersymmetric String Theories}. The resulting nine-dimensional model should have a circle radius $R$ as a parameter, which can be adjusted freely. Because we are considering closed string models, this nine-dimensional model should produce a ten-dimensional model (called model $M_2$) in $R\to0$ limit as well due to T-duality\cite{Casimir Effects in Superstring Theories,Vacuum Energies of String Compactified on Torus}. In particular, if model $M_1$ is supersymmetric and the $\boldsymbol{Z}_2$ action contains $(-1)^F$ where $F$ is the spacetime fermion number, the compactification causes SUSY breaking and the nine-dimensional interpolating model and model $M_2$ become non-supersymmetric. 

In Ref. \cite{Supersymmetry Restoration in the Compactified O(16) x O(16)-prime Heterotic String Theory,Small Cosmological Constant in String Models,Exponential suppression of the cosmological constant in nonsupersymmetric string vacua at two loops and beyond,Tension Between a Vanishing Cosmological Constant and Non-Supersymmetric Heterotic Orbifolds}, it is shown that in the near supersymmetric region of moduli space, the cosmological constant $\Lambda_{10}$ is written as follows:
\begin{equation}
\Lambda_{10} \simeq (N_F-N_B)\xi \tilde{a}^8 + \mathcal{O}(e^{-\tilde{a}^2}),\label{cc}
\end{equation} 
where $\xi$ is a positive constant and $\tilde{a}=a^{-1}=R/\sqrt{\alpha'}$, and $N_F$ ($N_B$) is the number of massless fermionic (bosonic) degrees of freedom.
Therefore, the cosmological constant is exponentially suppressed when $ N_F = N_B $. We would like to have non-supersymmetric models with $N_F=N_B$, but the nine-dimensional interpolating models with one parameter $R$ which we will review in section \ref{sec2} do not have such property no matter how one adjusts the parameter $R$. In order to generate cases with $N_F-N_B=0$, we need to increase the number of adjustable parameters. One such possibility is to compactify more dimensions. In this work, we instead consider nine-dimensional interpolating models with one more parameter by introducing a constant Wilson line background.


\section{Interpolating models with no Wilson line}\label{sec2}

In this section, we review the construction of an interpolating model which is originally proposed in Ref. \cite{Supersymmetry Restoration in the Compactified O(16) x O(16)-prime Heterotic String Theory}, and provide two concrete examples. In these examples, we provide the interpolations between the ten-dimensional non-supersymmetric $SO(16)\times SO(16)$ heterotic string model and one of the ten-dimensional supersymmetric heterotic strings\cite{The Heterotic String} as model $M_2$. The presentation below is based on Ref. \cite{Strong / weak coupling duality relations for nonsupersymmetric string theories,Towards a nonsupersymmetric string phenomenology}.\footnote{We can also construct these nine-dimensional string models by using free-fermionic construction\cite{Construction of Fermionic String Models in Four-Dimensions,Four-Dimensional Superstrings,Classification of Closed Fermionic String Models}.}

\subsection{The construction of interpolating models}\label{general construction}

Let us start from a flat ten-dimensional closed string model $M_1$ whose partition function is

\begin{equation}
\label{Z_M1}
Z_{M_1}=Z_{B}^{(8)}Z^{+}_{+},
\end{equation}
where $Z^{+}_{+}$ represents the contribution from the fermionic and the internal parts of string and $Z_{B}^{(n)}$ from the bosonic parts of string:
\begin{equation}
Z_{B}^{(n)}=\tau_2^{-n/2}\left(\eta \bar{\eta} \right)^{-n}.
\end{equation}
Let us first consider the circle compactification:
\begin{equation}
X^9\sim X^9+2\pi R.
\end{equation}
The left- and right-moving momenta along the compactified dimension are respectively
\begin{equation}\label{momenta}
p_L=\frac{1}{\sqrt{2\alpha'}}\left(na+\frac{w}{a} \right),~~ p_R=\frac{1}{\sqrt{2\alpha'}}\left(na-\frac{w}{a} \right),
\end{equation}
for $n,w\in \boldsymbol{Z}$. After the circle compactification, the partition function of model $M_1$ becomes
\begin{equation}
\label{circle}
Z^{(9)+}_{+}=\left( \left( \eta \bar{\eta}\right)^{-1} \sum_{n,w\in \boldsymbol{Z}} q^{\frac{\alpha'}{2} p_{L}^2} \bar{q}^{\frac{\alpha'}{2} p_{R}^2}\right) Z_{B}^{(7)}Z^{+}_{+}.
\end{equation}
In order to obtain two different ten-dimensional models at $R\to\infty$ and $R\to0$ limits, we have to consider the compactification on a twisted circle.
We choose $\mathcal{T}Q$ as the $\boldsymbol{Z}_2$ twist where $\mathcal{T}$ acts on the compactified circle as a half translation:
\begin{equation}
\label{half translation}
\mathcal{T}:\tilde{X}^9\to\tilde{X}^9+\pi \tilde{R}.
\end{equation}
Here, $\tilde{X}^9$ is the T-dualized coordinate for the compactified dimension and $\tilde{R}=\alpha'/R$ is the T-dualized radius.\footnote{It is not essential that a half translation $\mathcal{T}$ is accompanied with the T-dualized coordinate $\tilde{X}^9$. If we adopted the ordinary coordinate $X^9$, the sum in Eq. \eqref{lattices} would be over $n\in2(\boldsymbol{Z}+\alpha)$ and $w\in\boldsymbol{Z}+\beta$.} We denote by $Q$ a $\boldsymbol{Z}_2$ action that acts on the internal part of the string and that determines the two ten-dimensional models at the limits.

Because the $\boldsymbol{Z}_2$ twist contains $\mathcal{T}$, the partition function of the interpolating model contains a set of four momentum lattices:
\begin{equation}\label{lattices}
\begin{split}
\Lambda_{\alpha,\beta} &\equiv (\eta \bar{\eta})^{-1} \sum_{n\in\boldsymbol{Z}+\alpha,~w\in2(\boldsymbol{Z}+\beta)}q^{\frac{\alpha'}{2} p_{L}^2} \bar{q}^{\frac{\alpha'}{2} p_{R}^2}\\
&= (\eta \bar{\eta})^{-1} \sum_{n,w\in\boldsymbol{Z}}\exp\left[-\pi \left\lbrace\tau_2 \left(a^2(n+\alpha)^2+4a^{-2}(w+\beta)^2 \right)-4i\tau_1(n+\alpha)(w+\beta)  \right\rbrace  \right]. 
\end{split}
\end{equation}
where $\alpha$ and $\beta$ are 0 or 1/2, and $\alpha=0~(1/2)$ and $\beta=0~(1/2)$ imply the integer (half-integer) momenta and the even (odd) winding numbers respectively.
It is easy to show that under $\boldsymbol{T}:~\tau\to \tau+1$, $\Lambda_{\alpha,\beta}$ transforms as
\begin{equation}
\boldsymbol{T}:~\Lambda_{\alpha,\beta}\to e^{4\pi i \alpha \beta} \Lambda_{\alpha,\beta}.
\end{equation}
Under $\boldsymbol{S}:\tau\to -1/\tau$, by using the Poisson resummation formula, we obtain
\begin{equation}
\label{S変換}
\boldsymbol{S}:~\Lambda_{\alpha,\beta}\to \frac{1}{2}\sum_{\alpha',\beta'=0,1/2} e^{4\pi i(\alpha \beta'+\beta \alpha')}\Lambda_{\alpha',\beta'}.
\end{equation}
Note that, under $\boldsymbol{S}$ transformation, the combinations $\Lambda_{0,0}+\Lambda_{0,1/2}$ and $\Lambda_{1/2,0}-\Lambda_{1/2,1/2}$ are invariant and $\Lambda_{0,0}-\Lambda_{0,1/2}$ and $\Lambda_{1/2,0}+\Lambda_{1/2,1/2}$ are exchanged with each other.

Next, let us check the behaviors of $\Lambda_{\alpha,\beta}$ as $a\to 0~(R\to \infty)$ and as $a\to \infty~(R\to 0)$.
For $a\to 0$ limit, it is the part with zero coefficients of $a^{-2}$ in the exponential in Eq. \eqref{lattices} that give non-vanishing contributions. So only the lattices containing zero winding number are non-vanishing in the large $R$ limit:
\begin{equation}
\begin{split}
(\eta \bar{\eta})^{-1} \sum_{n\in \boldsymbol{Z}} \exp\left[ -\pi \left( a(n+\alpha)\right)^2 \right] \to (\eta \bar{\eta})^{-1} \int_{-\infty}^{\infty} \frac{dx}{a}e^{-\pi \tau_2 x^2}=\left( a \sqrt{\tau_2 }\eta \bar{\eta}\right) ^{-1},
\end{split}
\end{equation}
where $x=a(n+\alpha)$. Consequently, we see as $a\to0$
\begin{equation}
\label{a to 0}
\Lambda_{\alpha,0}\to \left( a \sqrt{\tau_2 }\eta \bar{\eta}\right) ^{-1},~~\Lambda_{\alpha,1/2}\to 0.
\end{equation}
On the other hand, in $a\to \infty$ limit, the non-vanishing contributions come from the lattices with zero momentum in Eq. \eqref{lattices}:
\begin{equation}
\begin{split}
(\eta \bar{\eta})^{-1} \sum_{w\in \boldsymbol{Z}} \exp\left[ -4\pi \left( \frac{w+\beta}{a}\right)^2 \right] \to (\eta \bar{\eta})^{-1} \int_{-\infty}^{\infty} \frac{dy}{a}e^{-4\pi \tau_2 y^2}=a\left( 2 \sqrt{\tau_2 }\eta \bar{\eta}\right) ^{-1},
\end{split}
\end{equation}
where $y=(w+\beta)/a$. Consequenly, we see as $a\to\infty$
\begin{equation}
\label{ainfinity}
\Lambda_{0,\beta}\to a\left( 2\sqrt{\tau_2 }\eta \bar{\eta}\right) ^{-1},~~\Lambda_{1/2,\beta}\to 0.
\end{equation}

Coming back to Eq. \eqref{circle}, we can rewrite as
\begin{equation}
Z^{(9)+}_{+}=\left( \Lambda_{0,0}+\Lambda_{0,1/2}\right) Z_{B}^{(7)}Z^{+}_{+},
\end{equation}
using $\Lambda_{\alpha,\beta}$. An interpolating model is obtained from $Z^{(9)+}_{+}$ by orbifolding with the $\boldsymbol{Z}_2$ action $\mathcal{T}Q$. A half translation $\mathcal{T}$ affects the lattices $\Lambda_{\alpha,\beta}$ and acts such that only the states with even winding numbers survive:
\begin{equation}
\mathcal{T}Q:~Z^{(9)+}_{+}\to Z^{(9)+}_{-} =\left( \Lambda_{0,0}-\Lambda_{0,1/2}\right) Z_{B}^{(7)}Z^{+}_{-},
\end{equation}
where $Z^{+}_{-}$ is defined as the $Q$-action of $Z^{+}_{+}$. The modular invariance requires the twisted sector\cite{Strings on Orbifolds,Strings on Orbifolds. 2.}. By using Eq. \eqref{S変換}, we see that under $\boldsymbol{S}$ transformation, $Z^{(9)+}_{-}$ transforms as
\begin{equation}
\boldsymbol{S}:~Z^{(9)+}_{-}\to Z^{(9)-}_{+}=\left( \Lambda_{1/2,0}+\Lambda_{1/2,1/2}\right) Z_{B}^{(7)}Z^{-}_{+},
\end{equation}
where $Z^{+}_{-}(-1/\tau)\equiv Z^{-}_{+}(\tau)$. Furthermore, when $\mathcal{T}Q$ acts on $Z^{(9)-}_{+}$, we obtain
\begin{equation}
\mathcal{T}Q:~Z^{(9)+}_{-}\to Z^{(9)-}_{-}=\left( \Lambda_{1/2,0}-\Lambda_{1/2,1/2}\right) Z_{B}^{(7)}Z^{-}_{-},
\end{equation}
where $Z^{-}_{-}$ is defined as the $Q$-action of $Z^{-}_{+}$. As a result, the total partition function which is modular invariant is
\begin{equation}
\label{Z_int}
\begin{split}
Z^{(9)}_{\text{int}}&=\frac{1}{2}\left(Z^{(9)+}_{+}+Z^{(9)+}_{-}+Z^{(9)-}_{+}+Z^{(9)-}_{-} \right) \\
&=\frac{1}{2}Z^{(7)}_{B}\left\lbrace \Lambda_{0,0}\left(Z^{+}_{+}+Z^{+}_{-}\right) +\Lambda_{0,1/2}\left(Z^{+}_{+}-Z^{+}_{-}\right)\right. \\ &\left. ~~~~~~~~+\Lambda_{1/2,0}\left(Z^{-}_{+}+Z^{-}_{-}\right)+\Lambda_{1/2,1/2}\left(Z^{-}_{+}-Z^{-}_{-}\right) \right\rbrace .
\end{split}
\end{equation}
In accordance with Eq. \eqref{ainfinity}, we see that $Z^{(9)}_{\text{int}}$ reproduces model $M_1$ in $a\to \infty$ limit. Note that the original model is reproduced as $R\to0$ as we have adopted the convention that a half translation $\mathcal{T}$ is introduced with regard to the T-dualized coordinate. If $\mathcal{T}$ were introduced with regard to the ordinary coordinate, the interpolating model would reproduce the original model $M_1$ in $R\to \infty$ limit. On the other hand, in $a\to 0$ limit, $Z^{(9)}_{\text{int}}$ produces model $M_2$ whose partition function is
\begin{equation}
Z_{M_2}=Z^{(8)}_{B}\left( Z^{+}_{+}+Z^{+}_{-}+Z^{-}_{+}+Z^{-}_{-}\right).
\end{equation}
That is, model $M_2$ is obtained by $Q$-twisting model $M_1$, which means that model $M_2$ is related to model $M_1$ by the $\boldsymbol{Z}_2$ action $Q$.

\subsection{Two examples} \label{two examples}
In this subsection, we review two examples of nine-dimensional interpolating models which are tachyon free for all radii.

As the first example, let us choose the ten-dimensional $SO(16)\times SO(16)$ heterotic model as model $M_1$ and the ten-dimensional supersymmetric $SO(32)$ heterotic model as model $M_2$:
\begin{align}
\label{so(16)so(16)}\nonumber
Z_{M_1}&=Z_{B}^{(8)}\left\lbrace \bar{O}_{8} \left(V_{16}C_{16}+C_{16}V_{16} \right)+\bar{V}_{8} \left(O_{16}O_{16}+S_{16}S_{16} \right)\right. \\ 
&~~~~~~~~~~~~\left. -\bar{S}_{8} \left(V_{16}V_{16}+C_{16}C_{16} \right)-\bar{C}_{8} \left(O_{16}S_{16}+S_{16}O_{16} \right)  \right\rbrace,\\
\label{so(32)}
Z_{M_2}&=Z_{B}^{(8)} \left( \bar{V}_{8}-\bar{S}_{8} \right)\left(O_{16} O_{16}+V_{16}V_{16}+S_{16}S_{16}+C_{16}C_{16}\right).
\end{align}
In this case, in the language of subsection \ref{general construction},
\begin{equation}
\begin{split}
Z^{+}_{+}&= \bar{O}_{8} \left(V_{16}C_{16}+C_{16}V_{16} \right)+\bar{V}_{8} \left(O_{16}O_{16}+S_{16}S_{16} \right)\\ 
&~~~~~~-\bar{S}_{8} \left(V_{16}V_{16}+C_{16}C_{16} \right)-\bar{C}_{8} \left(O_{16}S_{16}+S_{16}O_{16} \right) .  
\end{split}
\end{equation}
The $\boldsymbol{Z}_2$ action $Q$ which relates the $SO(16)\times SO(16)$ model to the supersymmetric $SO(32)$ model is $\bar{R}_{OC}$, which is defined as the reflection of the right-moving $SO(8)$ characters:
\begin{equation}
\begin{split}
\bar{R}_{OC}:&~\left( \bar{O}_{8},\bar{V}_{8},\bar{S}_{8},\bar{C}_{8} \right) \to \left( -\bar{O}_{8},\bar{V}_{8},\bar{S}_{8},-\bar{C}_{8} \right).\\
\end{split}
\end{equation}
Using this $\boldsymbol{Z}_2$ action $Q$ and the modular transformation of $SO(2n)$ characters
\begin{equation}
\boldsymbol{S}:~\left(
\begin{array}{c}
O_{2n}\\
V_{2n}\\
S_{2n}\\
C_{2n}\\
\end{array}
\right)
\to  \left(
\begin{array}{cccc}
1&1&1&1\\
1&1&-1&-1\\
1&-1&i^n&-i^n\\
1&-1&-i^n&i^n\\
\end{array}
\right)
\left(
\begin{array}{c}
O_{2n}\\
V_{2n}\\
S_{2n}\\
C_{2n}\\
\end{array}
\right),
\end{equation}
we have
\begin{equation}
\begin{split}
Z^{+}_{-}&=-\bar{O}_{8} \left(V_{16}C_{16}+C_{16}V_{16} \right)+\bar{V}_{8} \left(O_{16}O_{16}+S_{16}S_{16} \right)\\ 
&~~~~~~-\bar{S}_{8} \left(V_{16}V_{16}+C_{16}C_{16} \right)+\bar{C}_{8} \left(O_{16}S_{16}+S_{16}O_{16} \right),\\
Z^{-}_{+}&=\bar{O}_{8} \left(O_{16}S_{16}+S_{16}O_{16} \right)+\bar{V}_{8} \left(V_{16}V_{16}+C_{16}C_{16} \right)\\ 
&~~~~~~-\bar{S}_{8} \left(O_{16}O_{16}+S_{16}S_{16}\right)-\bar{C}_{8} \left(V_{16}C_{16}+C_{16}V_{16} \right) ,\\
Z^{-}_{-}&=-\bar{O}_{8} \left(O_{16}S_{16}+S_{16}O_{16} \right)+\bar{V}_{8} \left(V_{16}V_{16}+C_{16}C_{16} \right)\\ 
&~~~~~~-\bar{S}_{8} \left(O_{16}O_{16}+S_{16}S_{16}\right)+\bar{C}_{8} \left(V_{16}C_{16}+C_{16}V_{16} \right).
\end{split}
\end{equation}
Thus, from Eq. \eqref{Z_int}, we obtain the partition function of the interpolating model:
\begin{equation}
\label{example1}
\begin{split}
Z^{(9)}_{\text{int}}&= Z^{(7)}_{B}
\left\lbrace \Lambda_{0,0}\left(\bar{V}_{8}\left(O_{16}O_{16}+S_{16}S_{16} \right) -\bar{S}_{8}\left(V_{16}V_{16}+C_{16}C_{16} \right)\right) \right.  \\ 
&~~~~~~~~\left. +\Lambda_{0,1/2}\left(\bar{O}_{8}\left(V_{16}C_{16}+C_{16}V_{16} \right) -\bar{C}_{8}\left(O_{16}S_{16}+S_{16}O_{16} \right)\right) \right. \\
&~~~~~~~~\left. +\Lambda_{1/2,0}\left(\bar{V}_{8}\left(V_{16}V_{16}+C_{16}C_{16} \right) -\bar{S}_{8}\left(O_{16}O_{16}+S_{16}S_{16} \right)\right) \right.\\  
&~~~~~~~~\left. +\Lambda_{1/2,1/2}\left(\bar{O}_{8}\left(O_{16}S_{16}+S_{16}O_{16} \right) -\bar{C}_{8}\left(V_{16}C_{16}+C_{16}V_{16} \right)\right)  \right\rbrace .
\end{split}
\end{equation}
We can see that the first and the third lines of Eq. \eqref{example1} reproduce the non-supersymmetric $SO(16)\times SO(16)$ model \eqref{so(16)so(16)} while the first and the second lines the supersymmetric $SO(32)$ model \eqref{so(32)}. Note that this interpolating model is tachyon-free for a generic radius because there are no such terms as $\bar{O}_{8}O_{16}V_{16}$ or $\bar{O}_{8}V_{16}O_{16}$ in the partition function \eqref{example1}.

Let us see the massless spectrum of this model from the partition function \eqref{example1}. For a generic radius $0<R<\infty$, massless states can appear only when $n=w=0$, so we can find out the massless states by expanding the first line of Eq. \eqref{example1} in $q$. We list the expansion of each character in Appendix \ref{B1}. Then, for a generic radius, the massless spectrum of the model is
\begin{itemize}
	\item the nine-dimensional gravity multiplet: graviton $G_{\mu \nu}$, anti-symmetric tensor $B_{\mu \nu}$ and dilaton $\phi$;
	\item the gauge bosons transforming in the adjoint representation of $SO(16)\times SO(16)\times U(1)_{G,B}^{2}$;
	\item a spinor transforming in the ($\boldsymbol{16},\boldsymbol{16}$) of $SO(16)\times SO(16)$,
\end{itemize}
where $U(1)_{G,B}$ implies the Abelian factors generated by $G_{\mu 9}$ and $B_{\mu 9}$.
Note that this model has no points at which the gauge symmetry is enhanced in the region $0<R<\infty$.
Also, there are no points at which the cosmological constant is exponentially suppressed, that is, $N_F=N_B$, in all regions except $R\to \infty$. In $R\to \infty$ limit, the number of fermions is equal to that of bosons at each mass level including the massless level, which means that SUSY is restored in the limit.

In the second example, let us choose the $SO(16)\times SO(16)$ heterotic model as model $M_1$ and the supersymmetric $E_8\times E_8$ heterotic model as model $M_2$; $Z_{M_1}$ is the same as in the first example and
\begin{equation}
\label{E_8E_8}
Z_{M_2}=Z_{B}^{(8)}Z^{+}_{+}=Z_{B}^{(8)} \left( \bar{V}_{8}-\bar{S}_{8} \right)\left(O_{16}+S_{16} \right) \left( O_{16}+S_{16}\right).
\end{equation}
In this case, the $\boldsymbol{Z}_2$ action $Q$ is $R_{VC}$ which is defined as the reflection of one of the two left-moving $SO(16)$ characters: 
\begin{equation}
R_{VC}:~(O_{16},V_{16},S_{16},C_{16})\to(O_{16},-V_{16},S_{16},-C_{16}).
\end{equation}
The partition function of this interpolating model is obtained in a similar way to the first example:
\begin{equation}
\label{example2}
\begin{split}
Z^{(9)}_{\text{int}}&= Z^{(7)}_{B}
\left\lbrace \Lambda_{0,0}\left(\bar{V}_{8}\left(O_{16}O_{16}+S_{16}S_{16} \right) -\bar{S}_{8}\left(O_{16}S_{16}+S_{16}O_{16} \right)\right) \right.  \\ 
&~~~~~~~~\left. +\Lambda_{1/2,0}\left(\bar{V}_{8}\left(O_{16}S_{16}+S_{16}O_{16} \right) -\bar{S}_{8}\left(O_{16}O_{16}+S_{16}S_{16} \right)\right) \right.\\  
&~~~~~~~~\left. +\Lambda_{0,1/2}\left(\bar{O}_{8}\left(V_{16}C_{16}+C_{16}V_{16} \right) -\bar{C}_{8}\left(V_{16}V_{16}+C_{16}C_{16} \right)\right) \right. \\
&~~~~~~~~\left. +\Lambda_{1/2,1/2}\left(\bar{O}_{8}\left(V_{16}V_{16}+C_{16}C_{16} \right) -\bar{C}_{8}\left(V_{16}C_{16}+C_{16}V_{16} \right)\right)  \right\rbrace.
\end{split}
\end{equation}
For a generic radius $0<R<\infty$, the massless spectrum of this model is
\begin{itemize}
	\item the nine-dimensional gravity multiplet: graviton $G_{\mu \nu}$, anti-symmetric tensor $B_{\mu \nu}$ and dilaton $\phi$;
	\item the gauge bosons transforming in the adjoint representation of $SO(16)\times SO(16)\times U(1)_{G,B}^{2}$;
	\item a spinor transforming in the $(\boldsymbol{128},\boldsymbol{1})\oplus(\boldsymbol{1},\boldsymbol{128})$ of $SO(16)\times SO(16)$.
\end{itemize}
In this case, there are no points either where the gauge symmetry is enhanced or the cosmological constant is exponentially suppressed.

\section{Interpolating models with Wilson line}\label{sec.3}
The nine-dimensional interpolating models with the radius parameter $R$ in section \ref{sec2} do not give us a model with $N_F=N_B$ no matter how we adjust $R$. We need to increase the number of parameters in order to search for such a model. We can realize $N_F-N_B=0$ by compactifying more dimensions and adjusting the parameters of the compact manifold. For example, if the nine-dimensional model constructed in the previous example, in which $N_F-N_B=64$, are compactified on $(d-1)$-dimensional torus and the parameters of the torus are adjusted such that $ U(1)_{G,B}^{2d} $ is enhanced to $ U(1)_{G,B} ^{2d-r} \times G$, where $G$ is rank $r$ group which has eight non-zero roots, then we obtain interpolating models in which $N_F-N_B=0$. However, in this work, we will add one parameter by turning on Wilson line. In other words, we will generalize interpolating models by considering a twisted circle with a constant background. We expect that there are some conditions between parameters under which the gauge symmetry is enhanced as in Ref. \cite{New Heterotic String Theories in Uncompactified Dimensions $<$ 10,A Note on Toroidal Compactification of Heterotic String Theory,Toroidal Compactification of Nonsupersymmetric Heterotic Strings,A new twist on heterotic string compactifications}. In this section, we construct nine-dimensional interpolating models with two parameters by considering the compactification on a twisted circle with Wilson line.

Let us write the uncompactified dimensions as $X^{\mu}~(\mu=0,\cdots,9)$ and the internal ones as $X^{I}_{L}~(I=1,\cdots,16)$ for a ten-dimensional heterotic string model, and compactify the $X^9$-direction on a twisted circle. Furthermore, we switch on a constant Wilson line background with the components of $\mu=9$ and $I=1$ by adding to the worldsheet action
\begin{equation}\label{WL background}
A \int d^2 z \bar{\partial}X^{\mu=9}\partial X^{I=1}_{L}.
\end{equation}
It is only the momentum lattice of the center-of-mass mode that is affected by turning on Wilson line $A$. The addition of the constant Wilson line background corresponds to the boost on the momentum lattice \cite{New Heterotic String Theories in Uncompactified Dimensions $<$ 10,A Note on Toroidal Compactification of Heterotic String Theory,Eternal Higgs inflation and the cosmological constant problem}:
\begin{equation}
\left(
\begin{array}{c}
\ell_L  \\
p_L  \\
p_R 
\end{array}
\right)\to
\left(
\begin{array}{c}
\ell'_L  \\
p'_L  \\
p'_R 
\end{array}
\right)=R_{\ell_L\text{-}p_L}M_{\ell_L\text{-}p_R}\left(
\begin{array}{c}
\ell_L  \\
p_L  \\
p_R 
\end{array}
\right),
\end{equation}
where 
\begin{equation}\label{l_L}
 \ell_L=\frac{1}{\sqrt{\alpha'}}m
\end{equation}
is the left-moving momentum of the $X_{L}^{I=1}$-direction and $m\in\boldsymbol{Z}$ for the NS (anti-periodic) boundary condition and $m\in\boldsymbol{Z}+1/2$ for R (periodic). Here, $M_{\ell_L\text{-}p_R}$ and $R_{\ell_L\text{-}p_L}$ represent the boost on the $\ell_L$-$p_R$ plane and the rotation on the $\ell_L$-$p_L$ plane respectively. The boost $M_{\ell_L\text{-}p_R}$ is written in terms of $A$ as follows:
\begin{equation}
M_{\ell_L\text{-}p_R}=\left(
\begin{array}{ccc}
\sqrt{1+A^2}&0&A  \\
0&1&0  \\
A&0&\sqrt{1+A^2}
\end{array}
\right).
\end{equation}
We use $A$ to write $R_{\ell_L\text{-}p_L}$ as follows:
\begin{equation}
R_{\ell_L\text{-}p_L}=\left(
\begin{array}{ccc}
\frac{1}{\sqrt{1+A^2}}&-\frac{A}{\sqrt{1+A^2}}&0  \\
\frac{A}{\sqrt{1+A^2}}&\frac{1}{\sqrt{1+A^2}}&0  \\
0&0&1
\end{array}
\right).
\end{equation}
Therefore, after turning on Wilson line, we have
\begin{equation}
\label{p'}
\begin{split}
\ell'_L&=\frac{1}{\sqrt{2\alpha'}}\left( \sqrt{2}m-2\frac{A}{\sqrt{1+A^2}}\frac{w}{a} \right),\\
p'_L&=\frac{1}{\sqrt{2\alpha'}}\left( \sqrt{2}Am+\sqrt{1+A^2}an+\frac{1-A^2}{\sqrt{1+A^2}}\frac{w}{a} \right), \\
p'_L&=\frac{1}{\sqrt{2\alpha'}}\left( \sqrt{2}Am+\sqrt{1+A^2}an-\sqrt{1+A^2}\frac{w}{a} \right).
\end{split}
\end{equation}
The above equations mean that the left- and right-moving momenta of $X^{\mu=9}$ in Eq. \eqref{momenta} and the left-moving momentum of $X_{L}^{I=1}$ in Eq. \eqref{l_L} are mixed with each other by Wilson line. In terms of the functions in the partition function, the momentum lattice and a theta function in one of the two left-moving $SO(16)$ characters are convoluted as follows:
\begin{equation}
\label{変更点}
\Lambda_{\alpha,\beta}\left( \frac{\vartheta
	\begin{bmatrix} 
	\gamma\\ 
	\delta\\ 
	\end{bmatrix}}{\eta}\right) ^8 \to
\Lambda^{(\alpha,\beta)}_{(\gamma,\delta)}(a,A)\left( \frac{\vartheta
	\begin{bmatrix} 
	\gamma\\ 
	\delta\\ 
	\end{bmatrix}}{\eta}\right) ^7.
\end{equation}
Here, we define $\Lambda^{(\alpha,\beta)}_{(\gamma,\delta)}$ by
\begin{equation}
\label{mixed lattice}
\Lambda^{(\alpha,\beta)}_{(\gamma,\delta)}(a,A) \equiv \left( \eta \bar{\eta}\right)^{-1} \eta^{-1} \sum_{n,w,m} (-1)^{2m\delta} q^{\frac{\alpha'}{2}\left( p'^{2}_{L}+\ell'^{2}_{L}\right) } \bar{q}^{\frac{\alpha'}{2} p'^{2}_{R}},
\end{equation}
where the sum is taken over $n\in\boldsymbol{Z}+\alpha$,  $w\in2(\boldsymbol{Z}+\beta)$, $m\in\boldsymbol{Z}+\gamma$. Substituting Eq. \eqref{p'} into Eq. \eqref{mixed lattice}, we obtain
\begin{equation}
\Lambda^{(\alpha,\beta)}_{(\gamma,\delta)}(a,A)=\left( \eta \bar{\eta}\right)^{-1} \eta^{-1} \sum_{\boldsymbol{n}\in \boldsymbol{Z}^3} \exp\left[ -\pi \left( \boldsymbol{n}+\boldsymbol{x}\right)^{T} M(\tau_1,\tau_2;a,A) \left( \boldsymbol{n}+\boldsymbol{x}\right) + 2\pi i \boldsymbol{y}\cdot\boldsymbol{n} \right], 
\end{equation}
where $\boldsymbol{n}^T=(n,w,m)$, $\boldsymbol{x}^T=(\alpha,\beta,\gamma)$, $\boldsymbol{y}^T=(0,0,\delta)$ and $M(\tau_1,\tau_2;a,A)$ is a $3\times 3$ symmetric matrix of the following form:
\begin{equation}
M(\tau_1,\tau_2;a,A)=\left(
\begin{array}{ccc}
a^2 \sqrt{1+A^2} \tau_2&-2\left(A^2\tau_2+i\tau_1 \right) & \sqrt{2}aA\sqrt{1+A^2} \tau_2  \\
-2\left(A^2\tau_2+i\tau_1 \right)&4 a^{-2} \sqrt{1+A^2} \tau_2&-2\sqrt{2}a^{-1}A\sqrt{1+A^2} \tau_2  \\
\sqrt{2}aA\sqrt{1+A^2} \tau_2&-2\sqrt{2}a^{-1}A\sqrt{1+A^2} \tau_2& (1+2A^2)\tau_2-i\tau_1
\end{array}
\right).
\end{equation}
It is easy to see that, under $\boldsymbol{T}:~\tau\to\tau+1$,
\begin{equation}
\begin{split}
\Lambda^{(\alpha,\beta)}_{(0,\delta)}&\to e^{4\pi i \alpha\beta} \Lambda^{(\alpha,\beta)}_{(0,\delta+1/2)},\\
\Lambda^{(\alpha,\beta)}_{(1/2,\delta)}&\to e^{4\pi i \alpha\beta} e^{\pi i/4} \Lambda^{(\alpha,\beta)}_{(1/2,\delta+1/2)}
\end{split}
\end{equation}
Under $\boldsymbol{S}:~\tau\to-1/\tau$, by using the Poisson resummation formula, we obtain
\begin{equation}
\label{mixed lattice under S}
\Lambda^{(\alpha,\beta)}_{(\gamma,\delta)}\to \frac{1}{2} e^{2\pi i \gamma \delta}\sum_{\alpha',\beta'=0,1/2} e^{4\pi i \left(\alpha\beta'+\beta\alpha' \right) } \Lambda^{(\alpha',\beta')}_{(\delta,\gamma)}.
\end{equation} 

Before introducing some examples, let us discuss symmetry of the interpolating model. It is convenient to introduce a modular parameter $\tilde{\tau}$ in terms of the parameter of the twisted circle and Wilson line as
\begin{equation}
\tilde{\tau}=\tilde{\tau}_1+i\tilde{\tau}_2=\frac{A}{\sqrt{1+A^2}}\frac{1}{a}+i \frac{1}{\sqrt{1+A^2}}\frac{1}{a}.
\end{equation}
Note that $\left| \tilde{\tau}\right|^2 =1/a^2$, which means that the radial coordinate corresponds to radius $R$ and the angular coordinate to Wilson line $A$. Using $\tilde{\tau}$, momenta \eqref{p'} are rewritten as
\begin{equation}
\label{rewritten p'}
\begin{split}
\ell'_L&=\frac{1}{\sqrt{2\alpha'}}\left( \sqrt{2}m-2\tilde{\tau}_1 w \right),\\
p'_L&=\frac{1}{\sqrt{2\alpha'}}\frac{1}{\tilde{\tau}}_2\left( \sqrt{2}\tilde{\tau}_1m+n-(\tilde{\tau}_{1}^2-\tilde{\tau}_{2}^2)w \right), \\
p'_L&=\frac{1}{\sqrt{2\alpha'}}\frac{1}{\tilde{\tau}_2}\left( \sqrt{2}\tilde{\tau}_1m+n-(\tilde{\tau}_{1}^2+\tilde{\tau}_{2}^2)w \right),
\end{split}
\end{equation}
for $m\in\boldsymbol{Z}+\gamma$, $n\in\boldsymbol{Z}+\alpha$, $w\in2( \boldsymbol{Z}+\beta)$. From these momenta \eqref{rewritten p'}, we can see that the lattice $\Lambda^{(\alpha,\beta)}_{(\gamma,\delta)}$ is invariant under the shift
\begin{equation}
\label{shift}
\tilde{\tau}\to \tilde{\tau}+\sqrt{2}
\end{equation}
with the redefinitions
\begin{equation}
m\to m'=m-2w,~~~n\to n'=n+2m-2w,~~~w\to w'=w.
\end{equation}
Therefore the fundamental region of moduli space is\footnote{If the $\boldsymbol{Z}_2$ twist $\mathcal{T}Q$ acted trivially, then $n$ and $w$ would be both integers. Then, in addition to the shift \eqref{shift}, the momentum lattices would be invariant under $\tilde{\tau} \to -1/\tilde{\tau}$ with the replacement $n \leftrightarrow w$. This transformation would correspond to T-dual transformation, so the two limiting ten-dimensional models would be the same and the fundamental region would become $-\sqrt{2}/2\leq\tilde{\tau}_1\leq\sqrt{2}/2$ and $\left| \tilde{\tau} \right| \geq 1 $.}
\begin{equation}
\label{fund region}
-\frac{\sqrt{2}}{2}\leq\tilde{\tau}_1\leq\frac{\sqrt{2}}{2}.
\end{equation}

\subsection{The interpolation between SUSY $SO(32)$ and $SO(16)\times SO(16)$}\label{first example with WL}

As an example, let us include Wilson line in the first example of subsection \ref{two examples}.
According to Eq. \eqref{変更点}, the circle compactification of the $SO(16)\times SO(16)$ heterotic model with  Wilson line is
\begin{equation}
\label{so16 with WL}
\begin{split}
Z^{(9)}_{SO(16)\times SO(16)}(a,A)&=Z^{(9)+}_{+}(a,A)\\
&= Z_{B}^{(7)}\sum_{\beta=0,1/2}\left\lbrace 
\bar{O}_{8} \left( V_{16}^{(0,\beta)}(a,A) C_{16} +C_{16}^{(0,\beta)}(a,A) V_{16} \right)\right. \\
&~~~~~~~~~~~~~~~~~~\left. +\bar{V}_{8} \left( O_{16}^{(0,\beta)}(a,A) O_{16}+ S_{16}^{(0,\beta)}(a,A)S_{16} \right)\right. \\ 
&~~~~~~~~~~~~~~~~~~\left. -\bar{S}_{8} \left( V_{16}^{(0,\beta)}(a,A)V_{16} + C_{16}^{(0,\beta)}(a,A)C_{16} \right)\right. \\
&~~~~~~~~~~~~~~~~~~\left. -\bar{C}_{8} \left( O_{16}^{(0,\beta)}(a,A)S_{16} + S_{16}^{(0,\beta)}(a,A)O_{16} \right)  \right\rbrace,
\end{split}
\end{equation}
where $O^{(\alpha,\beta)}_{2n}$, $V^{(\alpha,\beta)}_{2n}$, $S^{(\alpha,\beta)}_{2n}$, $C^{(\alpha,\beta)}_{2n}$ are defined by
\begin{equation}\label{boosted character}
\begin{split}
O^{(\alpha,\beta)}_{2n}(a,A) &\equiv \frac{1}{2 \eta^{n-1}}\left( \Lambda^{(\alpha,\beta)}_{(0,0)}(a,A) \vartheta_{3}^{n-1} + \Lambda^{(\alpha,\beta)}_{(0,1/2)}(a,A) \vartheta_{4}^{n-1}\right), \\
V^{(\alpha,\beta)}_{2n}(a,A) &\equiv \frac{1}{2 \eta^{n-1}}\left( \Lambda^{(\alpha,\beta)}_{(0,0)}(a,A) \vartheta_{3}^{n-1} -\Lambda^{(\alpha,\beta)}_{(0,1/2)}(a,A) \vartheta_{4}^{n-1}\right), \\
S^{(\alpha,\beta)}_{2n}(a,A) &\equiv \frac{1}{2 \eta^{n-1}}\left( \Lambda^{(\alpha,\beta)}_{(1/2,0)}(a,A) \vartheta_{2}^{n-1} +\Lambda^{(\alpha,\beta)}_{(1/2,1/2)}(a,A) \vartheta_{1}^{n-1}\right), \\
C^{(\alpha,\beta)}_{2n}(a,A) &\equiv \frac{1}{2 \eta^{n-1}}\left( \Lambda^{(\alpha,\beta)}_{(1/2,0)}(a,A) \vartheta_{2}^{n-1} -\Lambda^{(\alpha,\beta)}_{(1/2,1/2)}(a,A) \vartheta_{1}^{n-1}\right). 
\end{split}
\end{equation}
We will refer to $O^{(\alpha,\beta)}_{n}$, $V^{(\alpha,\beta)}_{n}$, $S^{(\alpha,\beta)}_{n}$, $C^{(\alpha,\beta)}_{n}$ as boosted characters. In analogy with section \ref{sec2}, the interpolating model can be constructed from Eq. \eqref{so16 with WL} by orbifolding with the $\boldsymbol{Z}_2$ twist $\mathcal{T}Q$. In this case, $Q=\bar{R}_{OC}$ and the $\mathcal{T}$ action on the boosted characters changes an overall sign for $\beta=1/2$. Using Eq. \eqref{mixed lattice under S}, we find that under a $\boldsymbol{S}$ transformation, the boosted characters transform as
\begin{equation}
\left(
\begin{array}{c}
O^{(\alpha,\beta)}_{2n} \\
V^{(\alpha,\beta)}_{2n} \\
S^{(\alpha,\beta)}_{2n} \\
C^{(\alpha,\beta)}_{2n}
\end{array}
\right) \to \frac{1}{2}\sum_{\alpha',\beta'=0,1/2} e^{4\pi i (\alpha \beta'+\beta \alpha')} \left(
\begin{array}{cccc}
1&1&1&1 \\
1&1&-1&-1 \\
1&-1&i^n&-i^n \\
1&-1&-i^n&i^n
\end{array}
\right)\left(
\begin{array}{c}
O^{(\alpha',\beta')}_{2n} \\
V^{(\alpha',\beta')}_{2n} \\
S^{(\alpha',\beta')}_{2n} \\
C^{(\alpha',\beta')}_{2n}
\end{array}
\right).
\end{equation}
We obtain
\begin{equation}
\begin{split}
Z^{(9)+}_{-}&= Z_{B}^{(7)}\sum_{\beta=0,1/2}e^{2\pi i \beta}\left\lbrace 
-\bar{O}_{8} \left( V_{16}^{(0,\beta)} C_{16} +C_{16}^{(0,\beta)} V_{16} \right)
+\bar{V}_{8} \left( O_{16}^{(0,\beta)} O_{16}+ S_{16}^{(0,\beta)}S_{16} \right)\right. \\ 
&~~~~~~~~~~~~~~~~~~~~~~~~~~~
\left. -\bar{S}_{8} \left( V_{16}^{(0,\beta)}V_{16} + C_{16}^{(0,\beta)}C_{16} \right) +\bar{C}_{8} \left( O_{16}^{(0,\beta)}S_{16} + S_{16}^{(0,\beta)}O_{16} \right)  \right\rbrace,\\
Z^{(9)-}_{+}&= Z_{B}^{(7)}\sum_{\beta=0,1/2}\left\lbrace 
\bar{O}_{8} \left( O_{16}^{(1/2,\beta)} S_{16} +S_{16}^{(1/2,\beta)} O_{16} \right)
+\bar{V}_{8} \left( V_{16}^{(1/2,\beta)} V_{16}+ C_{16}^{(1/2,\beta)}C_{16} \right)\right. \\ 
&~~~~~~~~~~~~~~~~~~
\left. -\bar{S}_{8} \left( O_{16}^{(1/2,\beta)}O_{16} + S_{16}^{(1/2,\beta)}S_{16} \right) -\bar{C}_{8} \left( V_{16}^{(1/2,\beta)}C_{16} + C_{16}^{(1/2,\beta)}V_{16} \right)  \right\rbrace,\\
Z^{(9)-}_{-}&= Z_{B}^{(7)}\sum_{\beta=0,1/2}e^{2\pi i \beta}\left\lbrace 
-\bar{O}_{8} \left( O_{16}^{(1/2,\beta)} S_{16} +S_{16}^{(1/2,\beta)} O_{16} \right)
+\bar{V}_{8} \left( V_{16}^{(1/2,\beta)} V_{16}+ C_{16}^{(1/2,\beta)}C_{16} \right)\right. \\ 
&~~~~~~~~~~~~~~~~~~~~~~~~~~~
\left. -\bar{S}_{8} \left( O_{16}^{(1/2,\beta)}O_{16} + S_{16}^{(1/2,\beta)}S_{16} \right) +\bar{C}_{8} \left( V_{16}^{(1/2,\beta)}C_{16} + C_{16}^{(1/2,\beta)}V_{16} \right)  \right\rbrace.
\end{split}
\end{equation}
As a result of these equations, we find the total partition function of the interpolating model:
\begin{equation}
\label{example1 with WL}
\begin{split}
Z^{(9)}_{\text{int}}(a,A)&=\frac{1}{2} Z^{(7)}_{B} \left( Z^{(9)+}_{+}+Z^{(9)+}_{-}+Z^{(9)-}_{+}+Z^{(9)-}_{-} \right) \\
&= Z^{(7)}_{B}\left\lbrace \bar{V}_{8}\left( O^{(0,0)}_{16}O_{16}+S^{(0,0)}_{16}S_{16}\right) -\bar{S}_{8}\left( V^{(0,0)}_{16}V_{16}+C^{(0,0)}_{16}C_{16}\right)\right. \\
&~~~~~\left. +\bar{O}_{8}\left( V^{(0,1/2)}_{16}C_{16}+C^{(0,1/2)}_{16}V_{16}\right) -\bar{C}_{8}\left( O^{(0,1/2)}_{16}S_{16}+S^{(0,1/2)}_{16}O_{16}\right)\right.\\
&~~~~~\left. +\bar{V}_{8}\left( V^{(1/2,0)}_{16}V_{16}+C^{(1/2,0)}_{16}C_{16}\right) -\bar{S}_{8}\left( O^{(1/2,0)}_{16}O_{16}+S^{(1/2,0)}_{16}S_{16}\right)\right.\\
&~~~~~\left. +\bar{O}_{8}\left( O^{(1/2,1/2)}_{16}S_{16}+S^{(1/2,1/2)}_{16}O_{16}\right) -\bar{C}_{8}\left( V^{(1/2,1/2)}_{16}C_{16}+C^{(1/2,1/2)}_{16}V_{16}\right)\right\rbrace. \\
\end{split}
\end{equation}
Note that the only difference between Eq. \eqref{example1} and Eq. \eqref{example1 with WL} is that the momentum lattices are mixed with one of the two left-moving $SO(16)$ characters. Of course, it is easy to check that Eq. \eqref{example1 with WL} is equal to Eq. \eqref{example1} when $A=0$.

\subsubsection{The limiting cases}

Next, let us see the limiting cases $a\to0$ and $a\to \infty$ of the interpolating model \eqref{example1 with WL}. In the partition function \eqref{example1 with WL}, only the momentum lattices \eqref{mixed lattice} depend on $a$, so we need to see the behavior of $\Lambda^{(\alpha,\beta)}_{(\gamma,\delta)}$ in these limiting cases. As in the cases without Wilson line, the non-vanishing contributions come from the parts with zero winding number (momentum) in $a\to0$ ($a\to\infty$) limit, and $\Lambda^{(\alpha,1/2)}_{(\gamma,\delta)}$ ($\Lambda^{(1/2,\beta)}_{(\gamma,\delta)}$) vanishes as $a\to0$ ($a\to\infty$). As $a\to0$, we find
\begin{equation}
\label{endpt limit1}
\begin{split}
\Lambda^{(\alpha,0)}_{(\gamma,\delta)}(a,A) &\underset{w=0}{\simeq} (\eta\bar{\eta})^{-1}\eta^{-1}\sum_{n,m\in \boldsymbol{Z}}q^{(m+\gamma)^2/2}e^{2\pi im\delta} \\
&~~~~~~~~~~~~~\times \exp\left[ -\pi \tau_2 (1+A^2)\left( a(n+\alpha)+ \sqrt{2}\frac{A}{\sqrt{1+A^2}}(m+\gamma)\right)^2 \right]\\
&\to (\eta\bar{\eta})^{-1}\eta^{-1}\sum_{m\in \boldsymbol{Z}}q^{(m+\gamma)^2/2}e^{2\pi im\delta}\int_{-\infty}^{\infty}\frac{dx}{a}e^{-\pi \tau_2(1+A^2)x^2}\\
&=\frac{R_{\infty}}{\sqrt{\alpha'\tau_2}}(\eta\bar{\eta})^{-1}\eta^{-1}\vartheta
\begin{bmatrix} 
\gamma\\ 
\delta\\ 
\end{bmatrix},
\end{split}
\end{equation}
where $x\equiv a(n+\alpha)+ \sqrt{2}A(m+\gamma)/\sqrt{1+A^2}$ and $R_{\infty}\equiv R/\sqrt{1+A^2}$.  Similarly as $a\to\infty$, we find
\begin{equation}
\label{endpt limit2}
\begin{split}
\Lambda^{(0,\beta)}_{(\gamma,\delta)}(a,A) &\underset{n=0}{\simeq} (\eta\bar{\eta})^{-1}\eta^{-1}\sum_{w,m\in \boldsymbol{Z}}q^{(m+\gamma)^2/2}e^{2\pi im\delta} \\
&~~~~~~~~~~~~~\times \exp\left[ -4\pi \tau_2 (1+A^2)\left( \frac{w+\alpha}{a}- \frac{1}{\sqrt{2}}\frac{A}{\sqrt{1+A^2}}(m+\gamma)\right)^2 \right]\\
&\to (\eta\bar{\eta})^{-1}\eta^{-1}\sum_{m\in \boldsymbol{Z}}q^{(m+\gamma)^2/2}e^{2\pi im\delta}a\int_{-\infty}^{\infty}dye^{-4\pi \tau_2(1+A^2)y^2}\\
&=\frac{\sqrt{\alpha'}}{2\sqrt{\tau_2}R_{0}}(\eta\bar{\eta})^{-1}\eta^{-1}\vartheta
\begin{bmatrix} 
\gamma\\ 
\delta\\ 
\end{bmatrix},
\end{split}
\end{equation}
where $y\equiv  (w+\alpha)/a- {A}(m+\gamma)/{\sqrt{2(1+A^2)}}$ and $R_{0}\equiv \sqrt{1+A^2}R$. Note that $R_{\infty}$ ($R_{0}$) is the physical radius at the large (small) $R$ region. In fact, from Eq. \eqref{p'} we see
\begin{equation}
\begin{split}
\left.\left(  \ell'^{2}_{L}+p'^{2}_{L}\right) \right|_{m=w=0}&=\left. p'^{2}_{R}\right|_{m=w=0}=\frac{1}{2}\left( \frac{n}{R_{\infty}}\right) ^2,\\
\left.\left(  \ell'^{2}_{L}+p'^{2}_{L}\right) \right|_{m=n=0}&=\left. p'^{2}_{R}\right|_{m=n=0}=\frac{1}{2}\left( \frac{wR_{0}}{\alpha'}\right) ^2.
\end{split}
\end{equation}
Note that the effect of Wilson line is found only with the physical radii in the limiting cases.
In terms of the boosted characters, Eq. \eqref{endpt limit1} and Eq. \eqref{endpt limit2} respectively imply
\begin{equation}
\label{endpt limit3}
\begin{split}
\left( O_{n},V_{n},S_{n},C_{n}\right)^{(\alpha,\beta)}&\to \frac{R_{\infty}}{\sqrt{\alpha'\tau_2}}(\eta\bar{\eta})^{-1}\eta^{-1} \left( O_{n},V_{n},S_{n},C_{n}\right)\delta_{\beta,0}~~~~~(a\to0),\\
\left( O_{n},V_{n},S_{n},C_{n}\right)^{(\alpha,\beta)}&\to \frac{\sqrt{\alpha'}}{2\sqrt{\tau_2}R_{0}}(\eta\bar{\eta})^{-1}\eta^{-1} \left( O_{n},V_{n},S_{n},C_{n}\right)\delta_{\alpha,0}~~~~~(a\to\infty).
\end{split}
\end{equation}
Thus, Eq. \eqref{endpt limit3} shows that the interpolating model \eqref{example1 with WL} provides the $SO(16)\times SO(16)$ model at $a\to0$ and the supersymmetric $SO(32)$ model at $a\to \infty$ for any value of Wilson line $A$. 

\subsubsection{The massless spectrum}
Let us see the massless spectrum of this interpolating model for a generic set of values of $a$ and $A$. As is done in section \ref{sec2}, we can find out massless states from the parts with zero momentum and zero winding number of the partition function \eqref{example1 with WL}. By expanding the characters in $q$,\footnote{We list the expansion of the boosted characters \eqref{boosted character} in $q$ in Appendix \ref{B2}.} we find the following massless states for a generic set of values of $a$ and $A$:
\begin{itemize}
	\item the nine-dimensional gravity multiplet: graviton $G_{\mu \nu}$, anti-symmetric tensor $B_{\mu \nu}$ and dilaton $\phi$;
	\item the gauge bosons transforming in the adjoint representation of $SO(16)\times SO(14)\times U(1)\times U(1)_{G,B}^{2}$;
	\item a spinor transforming in the $(\boldsymbol{16},\boldsymbol{14})$ of $SO(16)\times SO(14)$.
\end{itemize}
Note that, compared to the first example in subsection \ref{two examples}, the gauge symmetry is broken to $SO(16)\times SO(14)\times U(1)$ because of Wilson line, and $N_F-N_B=32$. 

There are some conditions under which the additional massless states appear:
\begin{enumerate}
	\renewcommand{\theenumi}{(\Roman{enumi})}
	\renewcommand{\labelenumi}{\theenumi}
	\item $\tilde{\tau}_1=n_1/\sqrt{2}~~~~$  ($n_1\in\boldsymbol{Z}$)\label{case1-1}

	Using $a$ and $A$, this condition is rewritten as 
	\begin{equation}
	\sqrt{2}A+\sqrt{1+A^2}an_1=0,
	\end{equation}
	for any integer $n_1$. Under this condition, we find that the following additional massless states appear:
	\begin{itemize}
		\item two vectors transforming in the $(\boldsymbol{1},\boldsymbol{14})$ of $SO(16)\times SO(14)$; 
		\item two spinors transforming in the $(\boldsymbol{16},\boldsymbol{1})$ of $SO(16)\times SO(14)$.
	\end{itemize}
	These massless vectors and spinors come from $\bar{V}_{8} O^{(0,0)}_{16}O_{16}$ and $\bar{S}_{8} V^{(0,0)}_{16}V_{16}$ respectively when $(m,n)=(\pm1,\pm n_1)$ and $w=0$. This condition \ref{case1-1} thus enhances the gauge symmetry to $SO(16)\times SO(16)\times U(1)_{G,B}^{2}$, and at the same time, the massless spinor is promoted to transform in the $(\boldsymbol{16},\boldsymbol{16})$ of $SO(16)\times SO(16)$ as well. In this case, the additional massless fermionic and bosonic degrees of freedom are 256 and 224 respectively, and $N_F-N_B=64$.
	
	Note that condition \ref{case1-1} does not mean an infinite number of gauge enhanced orbits on the $\tilde{\tau}$-plane.
	Recalling the fundamental region \eqref{fund region} of the interpolating model, condition \ref{case1-1} implies that there are only two inequivalent $SO(16)\times SO(16)$ orbits. One of them is the $n_1=0$ orbit which corresponds to the case $A=0$. Thus, this orbit reproduces the first example in subsection \ref{two examples}.
	The other is the $n_1=1~(n_1=-1)$ orbit which is the new one that does not appear before considering the constant Wilson line background.
	
	\begin{figure}[t]\label{f1}
		\vspace{-0.0cm}
		\centering  
		\includegraphics[clip,width = 7.0cm]{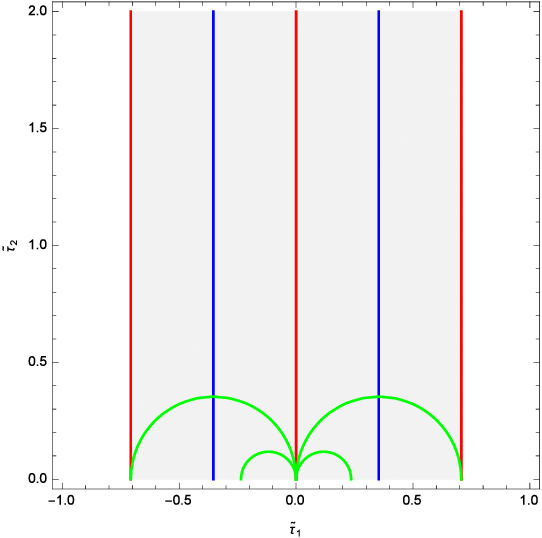}
		\caption[gauge enhanced orbit]{The shaded region is the fundamental region \eqref{fund region} and we plot the orbits on which the additional massless states appear in the first example. The three red lines correspond to condition \ref{case1-1} under which the gauge symmetry is enhanced to $SO(16)\times SO(16)$, and the one in the center implies the case of $A=0$. The two blue lines correspond to condition \ref{case1-2} under which the gauge symmetry is enhanced to $SO(18)\times SO(14)$. The green semi-circles correspond to condition \ref{case1-3} and we plot four orbits with $w_3=\pm1,\pm3$. }
		\vspace{0.0cm}
	\end{figure}
	
	\item $\tilde{\tau}_1=n_2/\sqrt{2}~~~~$  ($n_2\in\boldsymbol{Z}+1/2$)\label{case1-2}
	
	Under this condition, we find that the following additional massless states appear:
	\begin{itemize}
		\item two vectors transforming in the $(\boldsymbol{16},\boldsymbol{1})$ of $SO(16)\times SO(14)$; 
		\item two spinors transforming in the $(\boldsymbol{1},\boldsymbol{14})$ of $SO(16)\times SO(14)$.
	\end{itemize}
	These massless vectors and spinors come from $\bar{V}_{8} V^{(1/2,0)}_{16}V_{16}$ and $\bar{S}_{8} O^{(1/2,0)}_{16}O_{16}$ respectively when $(m,n)=(\pm1,\pm n_2)$ and $w=0$. This condition \ref{case1-2} thus enhances the gauge symmetry to $SO(18)\times SO(14)\times U(1)_{G,B}^{2}$, and at the same time, the massless spinor is promoted to transform in the $(\boldsymbol{18},\boldsymbol{14})$ of $SO(18)\times SO(14)$ as well. In this case, the additional massless fermionic and bosonic degrees of freedom are 224 and 256 respectively, which means $N_F-N_B=0$. The cosmological constant is exponentially suppressed on these orbits.
	
	Note that there are only two inequivalent orbits on which condition \ref{case1-2} is satisfied. For any half-integer $n_2$, all orbits are related either to the one with $n_2=1/2$ or the one with $n_2=-1/2$ by the shift \eqref{shift}.  
	
	\item $\frac{1}{\sqrt{2}}\tilde{\tau}_1-\left( \tilde{\tau}_{1}^2+\tilde{\tau}_{2}^2 \right)w_3=0 ~~~~$  ($w_2\in2\boldsymbol{Z}+1$)\label{case1-3}
	
	Using $a$ and $A$, this condition is rewritten as
	\begin{equation}
	\frac{1}{\sqrt{2}}A-\sqrt{1+A^2}\frac{w_3}{a}=0,
	\end{equation}
	for any odd integer $w_3$. The additional massless states are
	\begin{itemize}
		\item two conjugate spinors transforming in the $(\boldsymbol{1},\boldsymbol{64})$ of $SO(16)\times SO(14)$.
	\end{itemize}
	These massless conjugate spinors come from $\bar{C}_{8} S^{(0,0)}_{16}O_{16}$ when $(m,w)=\left( \pm1/2,\pm w_3\right) $ and $n=0$. Note that these conjugate spinors are the remnants of the $\boldsymbol{8}_{C}\otimes (\boldsymbol{1},\boldsymbol{128})$ in the ten-dimensional $SO(16)\times SO(16)$ model. 
\end{enumerate}

We plot these conditions in the fundamental region \eqref{fund region} of $\tilde{\tau}$-plane in Fig. 1. The Table 1 summarizes the conditions under which the additional massless states appear in this model. The table shows only the conditions with $w=0$ because we are interested in the large $R$ region where Eq. \eqref{cc} is valid.

\begin{table}[htb]
	\centering
	\begin{tabular}{|c||c|c|c|} \hline
		Conditions & $\tilde{\tau}_1=n_1/\sqrt{2}~$  ($n_1\in\boldsymbol{Z}$)& $\tilde{\tau}_1=n_2/\sqrt{2}~$  ($n_2\in\boldsymbol{Z}+1/2$) \\ \hline 
		Gauge group & $SO(16)\times SO(16)$& $SO(14) \times SO(18)$   \\ \hline
		$N_F-N_B$& positive & zero  \\ \hline
	\end{tabular}
\caption{The conditions}
\end{table}

\subsection{The interpolation between $E_8 \times E_8$ and $SO(16)\times SO(16)$}\label{second example with WL}

Next, let us include Wilson line in the second example of subsection \ref{two examples}. The starting point is the same as at subsection \ref{first example with WL} but the $Q$ action is $R_{VC}$ in this case. According to the construction in subsection \ref{general construction}, we find that the total partition function is
\begin{equation}
\label{example2 with WL}
\begin{split}
Z^{(9)}_{\text{int}}(a,A)&=\frac{1}{2} Z^{(7)}_{B} \left( Z^{(9)+}_{+}+Z^{(9)+}_{-}+Z^{(9)-}_{+}+Z^{(9)-}_{-} \right) \\
&= Z^{(7)}_{B}\left\lbrace \bar{V}_{8}\left( O^{(0,0)}_{16}O_{16}+S^{(0,0)}_{16}S_{16}\right) -\bar{S}_{8}\left( O^{(0,0)}_{16}S_{16}+S^{(0,0)}_{16}O_{16}\right)\right. \\
&~~~~~\left. +\bar{O}_{8}\left( V^{(0,1/2)}_{16}C_{16}+C^{(0,1/2)}_{16}V_{16}\right) -\bar{C}_{8}\left( V^{(0,1/2)}_{16}V_{16}+C^{(0,1/2)}_{16}C_{16}\right)\right.\\
&~~~~~\left. +\bar{V}_{8}\left( O^{(1/2,0)}_{16}S_{16}+S^{(1/2,0)}_{16}O_{16}\right) -\bar{S}_{8}\left( O^{(1/2,0)}_{16}O_{16}+S^{(1/2,0)}_{16}S_{16}\right)\right.\\
&~~~~~\left. +\bar{O}_{8}\left( V^{(1/2,1/2)}_{16}V_{16}+C^{(1/2,1/2)}_{16}C_{16}\right) -\bar{C}_{8}\left( V^{(1/2,1/2)}_{16}C_{16}+C^{(1/2,1/2)}_{16}V_{16}\right)\right\rbrace. \\
\end{split}
\end{equation}

Using the limiting behaviors of the boosted characters \eqref{endpt limit3}, we can see that this interpolating model \eqref{example2 with WL} reproduces the supersymmetric $E_8\times E_8$ model and the $SO(16)\times SO(16)$ model as $a\to 0$ and $a\to \infty$ respectively, for any value of $A$. 

\subsubsection{The massless spectrum}

Let us see the massless spectrum of this interpolating model for a generic set of values of $a$ and $A$. By expanding the partition function \eqref{example2 with WL} in $q$, we find
\begin{itemize}
	\item the nine-dimensional gravity multiplet: graviton $G_{\mu \nu}$, anti-symmetric tensor $B_{\mu \nu}$ and dilaton $\phi$;
	\item the gauge bosons transforming in the adjoint representation of $SO(16)\times SO(14)\times U(1)\times U(1)_{G,B}^{2}$;
	\item a spinor transforming in the $(\boldsymbol{128},\boldsymbol{1})$ of $SO(16)\times SO(14)$.
\end{itemize}
These massless states come from $\bar{V}_{8}O^{(0,0)}_{16}O_{16}$ or $\bar{S}_{8}O^{(0,0)}_{16}S_{16}$. For a generic set of values of $a$ and $A$, $N_F-N_B=-736$, and the cosmological constant becomes negative. We can find that there are some conditions between $a$ and $A$ under which the additional massless states appear:
\begin{enumerate}
	\renewcommand{\theenumi}{(\Roman{enumi})}
	\renewcommand{\labelenumi}{\theenumi}
	\item $\tilde{\tau}_1=n_1/\sqrt{2}~~~~$  ($n_1\in\boldsymbol{Z}$) \label{case2-1}
	
	Under this condition, we find that the following additional massless states appear:
	\begin{itemize}
		\item two vectors transforming in the $(\boldsymbol{1},\boldsymbol{14})$ of $SO(16)\times SO(14)$.
	\end{itemize}
	These massless vectors come from $\bar{V}_{8} O^{(0,0)}_{16}O_{16}$ when $(m,n)=(\pm1,\pm n_1)$ and $w=0$. This condition \ref{case2-1} thus enhances the gauge symmetry to $SO(16)\times SO(16)\times U(1)_{G,B}^{2}$. Furthermore, the different additional massless states appear depending on whether $n_1$ is even or odd:
	\makeatletter
	\renewcommand{\p@enumii}{}
	\makeatother
	\begin{enumerate}
		\renewcommand{\theenumii}{(I-\alph{enumii})}
		\renewcommand{\labelenumii}{\theenumii}
		\item $n_1\in 2\boldsymbol{Z}$\label{case2-1-a}
		\begin{itemize}
			\item two spinors transforming in the $(\boldsymbol{1},\boldsymbol{64})$ of $SO(16)\times SO(14)$.
		\end{itemize}
		These states come from $\bar{S}_{8} S^{(0,0)}_{16}O_{16}$ when $(m,n)=(\pm1/2,\pm n_1/2)$ and $w=0$. In the representation of the $SO(16)\times SO(16)$, this is a spinor transforming in the $(\boldsymbol{1},\boldsymbol{128})$. Note that in the fundamental region \eqref{fund region}, this condition corresponds to the $\tilde{\tau}_1=0$ orbit, which means the case $A=0$. The massless spectrum under this condition is thus the same as that of the second example in subsection \ref{two examples}.
		
		\item $n_1\in 2\boldsymbol{Z}+1$\label{case2-1-b}
		\begin{itemize}
			\item two vectors transforming in the $(\boldsymbol{1},\boldsymbol{64})$ of $SO(16)\times SO(14)$.
		\end{itemize}
		
		These states come from $\bar{V}_{8} S^{(1/2,0)}_{16}O_{16}$ when $(m,n)=(\pm1/2,\pm n_1/2)$ and $w=0$. In representation of the $SO(16)\times SO(16)$, this is a vector transforming in the $(\boldsymbol{1},\boldsymbol{128})$. Therefore, under this condition, the gauge symmetry is enhanced to $SO(16)\times E_8$ beyond $SO(16)\times SO(16)$. Note that in the fundamental region \eqref{fund region}, this condition corresponds to the $\tilde{\tau}_1=\sqrt{2}/2$ (or $\tilde{\tau}_1=-\sqrt{2}/2$) orbit.
	\end{enumerate}
	
	\begin{figure}[t]\label{f2}
		\centering  
		\includegraphics[clip,width = 7.0cm]{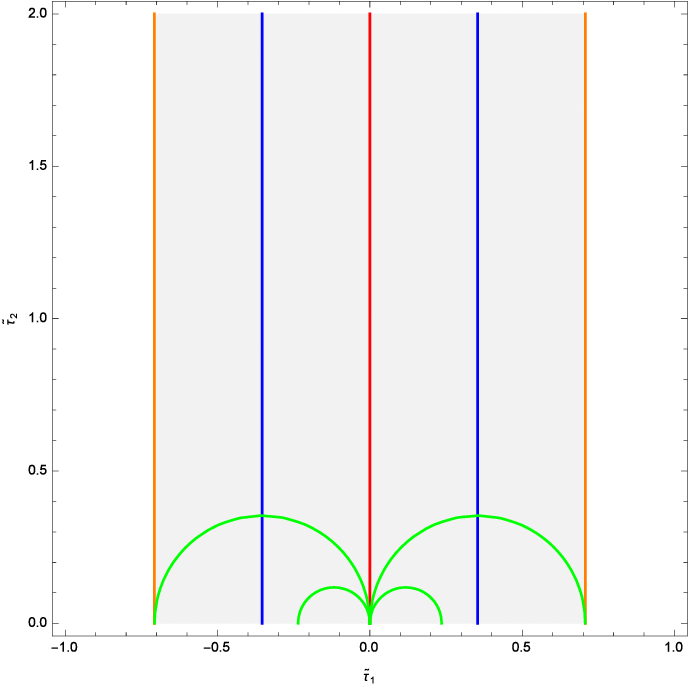}
		\caption[gauge enhanced orbit]{The shaded region is the fundamental region \eqref{fund region} and we plot the orbits on which additional massless states appear in the second example. The red line corresponds to condition \ref{case2-1-a} under which the gauge symmetry is enhanced to $SO(16)\times SO(16)$. The two orange lines correspond to condition \ref{case2-1-b} under which the gauge symmetry is enhanced to $SO(16)\times E_8$. The two blue lines correspond to condition \ref{case2-2}. The green semi-circles correspond to condition \ref{case2-3} and we plot four orbits with $w_3=\pm1,\pm3$.}
		\vspace{-0.0cm}
	\end{figure}
	
	\item $\tilde{\tau}_1=n_2/\sqrt{2}~~~~$  ($n_2\in\boldsymbol{Z}+1/2$)\label{case2-2}
	
	Under this condition, we find that the following additional massless states appear:
	\begin{itemize}
		\item two spinors transforming in the $(\boldsymbol{1},\boldsymbol{14})$ of $SO(16)\times SO(14)$. 
	\end{itemize}
	These massless spinors come from $\bar{S}_{8} O^{(1/2,0)}_{16}O_{16}$ when $(m,n)=(\pm1,\pm n_2)$ and $w=0$. 
	Note that in the fundamental region \eqref{fund region}, this condition corresponds to the two orbits which are $\tilde{\tau}_1=\sqrt{2}/4$ and $\tilde{\tau}_1=-\sqrt{2}/4$.

	\item $\frac{1}{\sqrt{2}}\tilde{\tau}_1-\left( \tilde{\tau}_{1}^2+\tilde{\tau}_{2}^2 \right)w_3=0 ~~~~$  ($w_3\in2\boldsymbol{Z}+1$)\label{case2-3}
	
	The additional massless states are
	\begin{itemize}
		\item two conjugate spinors transforming in the $(\boldsymbol{16},\boldsymbol{1})$ of $SO(16)\times SO(14)$.
	\end{itemize}
	These massless conjugate spinors come from $\bar{C}_{8} V^{(0,1/2)}_{16}V_{16}$ when $(m,w)=\left( \pm1/2,\pm w_3\right) $ and $n=0$. Note that these conjugate spinors are the remnants of the $\boldsymbol{8}_{C}\otimes (\boldsymbol{16},\boldsymbol{16})$ in the ten-dimensional $SO(16)\times SO(16)$ model. 
\end{enumerate}

We plot these conditions in the fundamental region \eqref{fund region} of $\tilde{\tau}$-plane in Fig. 2. The Table 2 summarizes the conditions under which the additional massless states appear in this model.

\begin{table}[htb]
	\begin{tabular}{|c||c|c|c|} \hline
		Conditions & $\tilde{\tau}_1=n_1/\sqrt{2}~$  ($n_1\in2\boldsymbol{Z}$)& $\tilde{\tau}_1=n_1/\sqrt{2}~$  ($n_1\in2\boldsymbol{Z}+1$)& $\tilde{\tau}_1=n_2/\sqrt{2}~$  ($n_2\in\boldsymbol{Z}+1/2$) \\ \hline 
		Gauge group & $SO(16)\times SO(16)$& $SO(16) \times E_8$   &$SO(16)\times SO(14)\times U(1)$\\ \hline
		$N_F-N_B$& positive & negative  & negative\\ \hline
	\end{tabular}
	\caption{The conditions}
\end{table}

Finally, let us mention that in these models considered in this section, it is straightforward to calculate tree and one-loop scattering amplitudes of massless particles to obtain signals of broken supersymmetry\cite{Itoyama:1987qz,Itoyama:2015fht,Arakane:1996rh,Kostelecky:1986xg}.
\section{Conclusions}
We have constructed nine-dimensional interpolating models with two parameters by considering the compactification on a twisted circle with the constant Wilson line background \eqref{WL background}, and have studied the massless spectra of these models. Furthermore, we have found some conditions between circle radius $R$ and Wilson line $A$ under which additional massless states are present.
In the nine-dimensional model that interpolates between the ten-dimensional supersymmetric $SO(32)$ model and the ten-dimensional $SO(16)\times SO(16)$ model, we find the conditions under which the gauge symmetry is enhanced to $SO(16)\times SO(16)$ or $SO(18)\times SO(14)$. Especially, under the second condition, the massless fermionic and bosonic degrees of freedom become equal, which means that the cosmological constant is exponentially suppressed. Recent references related to this point include \cite{Kounnas:2016gmz,Kounnas:2017mad,Florakis:2016ani}. According to Ref. \cite{Stability and vacuum energy in open string models with broken supersymmetry}, which is carried out in the type I dual picture\cite{Evidence for heterotic - type I string duality}, the brane configuration with the gauge group $SO(18)\times SO(14)$ yields the nine-dimensional non-supersymmetric model with $N_F-N_B=0$, although it has tachyonic directions in moduli space. 
On the other hand, our interpolation between the ten-dimensional supersymmetric $E_8\times E_8$ model and the ten-dimensional $SO(16)\times SO(16)$ model did not produce the condition with $N_F-N_B=0$. We have however found the conditions under which the gauge symmetry is enhanced to $SO(16)\times SO(16)$ or $SO(16)\times E_8$. 

As one of the future works, we have to discuss the stability of Wilson line as in Ref. \cite{Stability and vacuum energy in open string models with broken supersymmetry,Kounnas:2016gmz,Kounnas:2017mad,Florakis:2016ani}. Even if the cosmological constant is very small on a certain point (orbit) of moduli space, it is not clear that Wilson line is stable on the point (orbit). Namely, we need to write down the cosmological constant in terms of Wilson line and find the stable points of Wilson line.
\section*{Acknowledgments}
The work of H. I. was partially supported by JSPS KAKENHI Grant Number 19K03828.


\appendix

\section{Notation for the partition functions}

We summarize the notation for some functions that appear in the partition functions. The Dedekind eta function is 
\begin{equation}
\eta(\tau)=q^{-1/24}\prod_{n=1}^{\infty}\left( 1-q^{n}\right),
\end{equation} 
where $q=e^{2\pi i\tau}$. The theta function with characteristics is defined by
\begin{equation}
\vartheta
\begin{bmatrix} 
\alpha\\ 
\beta\\ 
\end{bmatrix}(z,\tau)=\sum_{n=-\infty}^{\infty}\exp\left( \pi i (n+\alpha)^2 \tau +2\pi i (n+\alpha)(z+\beta) \right). 
\end{equation}
Especially, when $\alpha$ and $\beta$ are 0 or 1/2 and $z=0$, we use the following shorthand notations:
\begin{align}
\vartheta_{1}(\tau)&=\vartheta
\begin{bmatrix} 
1/2\\ 
1/2\\ 
\end{bmatrix}(0,\tau)=0,\\
\vartheta_{2}(\tau)&=\vartheta
\begin{bmatrix} 
1/2\\ 
0\\ 
\end{bmatrix}(0,\tau),\\
\vartheta_{3}(\tau)&=\vartheta
\begin{bmatrix} 
0\\ 
0\\ 
\end{bmatrix}(0,\tau),\\
\vartheta_{4}(\tau)&=\vartheta
\begin{bmatrix} 
0\\ 
1/2\\ 
\end{bmatrix}(0,\tau).
\end{align}
These theta functions satisfy the Jacobi's abstruse identity:
\begin{equation}
\vartheta_{3}(\tau)^4-\vartheta_{4}(\tau)^4-\vartheta_{2}(\tau)^4=0.
\end{equation}
We write the $SO(2n)$ characters in terms of the theta functions as follows:
\begin{align}
O_{2n}&=\frac{1}{2\eta^{n}}\left( \vartheta_{3}^{n}+\vartheta_{4}^{n} \right) ,\\
V_{2n}&=\frac{1}{2\eta^{n}}\left( \vartheta_{3}^{n}-\vartheta_{4}^{n} \right) ,\\
S_{2n}&=\frac{1}{2\eta^{n}}\left( \vartheta_{2}^{n}+\vartheta_{1}^{n} \right) ,\\
C_{2n}&=\frac{1}{2\eta^{n}}\left( \vartheta_{2}^{n}-\vartheta_{1}^{n} \right).
\end{align}
In terms of the characters, the Jacobi's abstruse identity is
\begin{equation}
V_{8}-S_{8}=0.
\end{equation}

\section{The expansions of the characters}\label{characters expansion}

In string theories, we can see the spectrum of each mass levels by expanding the partition function in $q$. In this appendix, in order to see the massless states, which are the coefficients of $q^{0}$, we shall expand the $SO(8)$ and $SO(16)$ characters, which appear in the partition function of some heterotic models\footnote{There are five ten-dimensional heterotic models whose partition functions are expressed in terms of the characters $SO(8)$ or $SO(16)$: the supersymmetric $SO(32)$ model, the supersymmetric $E_{8}\times E_{8}$ model, the non-supersymmetric $SO(32)$ model, the $SO(16)\times E_8$ model, the $SO(16)\times SO(16)$ model.}.

\subsection{The case with no Wilson line}\label{B1}
For section \ref{sec2}, we expand $\eta^{-8}\left( O_{2n},V_{2n},S_{2n},C_{2n} \right)$ where $\eta^8$ is the contribution from $X^{m}$ and the $SO(2n)$ characters are from $\psi^{m}$ or $X_{L}^{I}$, where $m=2,\cdots,10$ and $I=1,\cdots,16$:
\begin{align}
\eta^{-8}O_{2n}&=q^{-8/24-n/24}\left( 1+\frac{2n(2n-1)}{2}q+8q+\mathcal{O}(q^2) \right),\label{O_8}\\
\eta^{-8}V_{2n}&=q^{-8/24-n/24+1/2}\left(2n+\mathcal{O}(q) \right),\label{V_8}\\
\eta^{-8}S_{2n}&=\eta^{-8}C_{2n}=q^{-8/24+n/12}\left(2^{n-1}+\mathcal{O}(q) \right).\label{S_8}
\end{align}
Note that the lowest order terms of \eqref{O_8}, \eqref{V_8} and \eqref{S_8} correspond to the degrees of freedom of the identity, the vector and the spinor (the conjugate spinor) respectively, and the second term of \eqref{O_8} to the adjoint representation of $SO(2n)$. 
The third term of $\eta^{-8}O_{2n}$ comes from $\eta^{-8}$, that is, the contributions from $X^{m}$. 

The right moving parts of the partition functions are expanded as
\begin{align}
\bar{\eta}^{-8}\bar{O}_{8}&=\bar{q}^{-1/2}\left( 1+\frac{2n(2n-1)}{2}\bar{q}+8\bar{q}+\mathcal{O}(\bar{q}^2) \right),\\
\bar{\eta}^{-8}\bar{V}_{8}&=8+\mathcal{O}(\bar{q}),\\
\bar{\eta}^{-8}\bar{S}_{8}&=\bar{\eta}^{-8}\bar{S}_{8}=8+\mathcal{O}(\bar{q}).
\end{align}
The left moving parts of the partition functions in some heterotic models might include
\begin{align}
\eta^{-8}O_{16}O_{16}&=q^{-1}\left(1+2\cdot\frac{16\cdot 15}{2}+8q+\mathcal{O}(q^2) \right),\\
\eta^{-8}O_{16}V_{16}&=q^{-1/2} \left(2n+\mathcal{O}(q) \right),\\
\eta^{-8}O_{16}S_{16}&=\eta^{-8}O_{16}C_{16}=2^{n-1}+\mathcal{O}(q),\\
\eta^{-8}V_{16}V_{16}&=16\cdot 16+\mathcal{O}(q),\\
\eta^{-8}V_{16}S_{16}&=\eta^{-8}V_{16}C_{16}=q^{-1/2} \left(2n\cdot 2^{n-1}+\mathcal{O}(q) \right),\\
\eta^{-8}S_{16}S_{16}&=\eta^{-8}S_{16}C_{16}=q\left( 2^{2(n-1)}+\mathcal{O}(q)\right).
\end{align}
Note that all states which come from $\eta^{-8}V_{16}S_{16}$ or $\eta^{-8}S_{16}S_{16}$ ($\eta^{-8}S_{16}C_{16}$) are massive, and tachyons can appear only from  the combination $\left( \eta\bar{\eta}\right)^{-8} \bar{O}_{8}O_{16}V_{16}$ because of the level-matching condition. 

\subsection{The case with Wilson line}\label{B2}
As section \ref{sec.3}, when Wilson line is switched on, the left-moving $SO(16)$ characters and the momentum lattices are mixed. So, in such a case, we need to expand the boosted characters \eqref{boosted character} in order to see the spectrum. The boosted characters are expanded as follows:

\begin{equation}
\begin{split}
O^{(\alpha,\beta)}_{16}&= \frac{1}{2 \eta^{7}}\left( \Lambda^{(\alpha,\beta)}_{(0,0)} \vartheta_{3}^{7} + \Lambda^{(\alpha,\beta)}_{(0,1/2)} \vartheta_{4}^{7}\right) \\
&=(\eta\bar{\eta})^{-1}q^{-\frac{8}{24}}\sum_{n,w}\left\lbrace \sum_{m\in 2\boldsymbol{Z}}q^{\frac{\alpha'}{2}\left(\ell'^{2}_{L}+p'^{2}_{L} \right)}\bar{q}^{\frac{\alpha'}{2}p'^{2}_{R}}  \left( 1+q+\frac{14\cdot 13}{2}q+\mathcal{O}(q^{\frac{3}{2}}) \right)\right. \\ 
&\left.~~~~~~~~~~~~~~~~~~~~~~~~~~~~~
+\sum_{m\in 2\boldsymbol{Z}+1}q^{\frac{\alpha'}{2}\left(\ell'^{2}_{L}+p'^{2}_{L} \right)} \bar{q}^{\frac{\alpha'}{2}p'^{2}_{R}}  \left( 14q^{\frac{1}{2}}+\mathcal{O}(q^{\frac{3}{2}}) \right) \right\rbrace, \\
V^{(\alpha,\beta)}_{16}&= \frac{1}{2 \eta^{7}}\left( \Lambda^{(\alpha,\beta)}_{(0,0)} \vartheta_{3}^{7} - \Lambda^{(\alpha,\beta)}_{(0,1/2)} \vartheta_{4}^{7}\right) \\
&=(\eta\bar{\eta})^{-1}q^{-\frac{8}{24}}\sum_{n,w}\left\lbrace \sum_{m\in 2\boldsymbol{Z}}q^{\frac{\alpha'}{2}\left(\ell'^{2}_{L}+p'^{2}_{L} \right)}\bar{q}^{\frac{\alpha'}{2}p'^{2}_{R}}  \left( 14q^{\frac{1}{2}}+\mathcal{O}(q^{\frac{3}{2}}) \right)\right. \\ 
&\left.~~~~~~~~~~~~~~~~~~
+\sum_{m\in 2\boldsymbol{Z}+1}q^{\frac{\alpha'}{2}\left(\ell'^{2}_{L}+p'^{2}_{L} \right)} \bar{q}^{\frac{\alpha'}{2}p'^{2}_{R}}  \left( 1+q+\frac{14\cdot 13}{2}q+\mathcal{O}(q^{\frac{3}{2}}) \right) \right\rbrace, \\
S^{(\alpha,\beta)}_{16}&=C^{(\alpha,\beta)}_{16}= \frac{1}{2 \eta^{7}}\left( \Lambda^{(\alpha,\beta)}_{(1/2,0)} \vartheta_{2}^{7} \pm \Lambda^{(\alpha,\beta)}_{(1/2,1/2)} \vartheta_{1}^{7}\right) \\
&=(\eta\bar{\eta})^{-1}q^{-\frac{1}{24}+\frac{7}{12}}\sum_{n,w}\left\lbrace \sum_{m\in \boldsymbol{Z}+1/2}q^{\frac{\alpha'}{2}\left(\ell'^{2}_{L}+p'^{2}_{L} \right)}\bar{q}^{\frac{\alpha'}{2}p'^{2}_{R}}  \left( 2^{7-1}+\mathcal{O}(q) \right) \right\rbrace, \\
\end{split}
\end{equation}
where the sum is taken over $n\in \boldsymbol{Z}+\alpha$ and $w\in 2(\boldsymbol{Z}+\beta)$. As we are interested only in the left-moving parts of the partition function, we expand the following products:

\begin{equation}
\begin{split}
&\bar{\eta}\eta^{-7}O^{(\alpha,\beta)}_{16}O_{16}
=q^{-1}\sum_{n,w}\left\lbrace \sum_{m\in 2\boldsymbol{Z}}q^{\frac{\alpha'}{2}\left(\ell'^{2}_{L}+p'^{2}_{L} \right)}\bar{q}^{\frac{\alpha'}{2}p'^{2}_{R}}  \left( 1+8q+\left(\frac{16\cdot 15}{2}+ \frac{14\cdot 13}{2}+1\right) q+\mathcal{O}(q^{\frac{3}{2}}) \right)\right. \\ 
&\left.~~~~~~~~~~~~~~~~~~~~~~~~~~~~~~~~~~~~~~~~~~~~~~~~~~~~~~~~~~~~~~
+\sum_{m\in 2\boldsymbol{Z}+1}q^{\frac{\alpha'}{2}\left(\ell'^{2}_{L}+p'^{2}_{L} \right)} \bar{q}^{\frac{\alpha'}{2}p'^{2}_{R}}  \left( 1\cdot 14q^{\frac{1}{2}}+\mathcal{O}(q^{\frac{3}{2}}) \right) \right\rbrace,\\
&\bar{\eta}\eta^{-7}O^{(\alpha,\beta)}_{16}V_{16}\\
&=q^{-\frac{1}{2}}\sum_{n,w}\left\lbrace \sum_{m\in 2\boldsymbol{Z}}q^{\frac{\alpha'}{2}\left(\ell'^{2}_{L}+p'^{2}_{L} \right)}\bar{q}^{\frac{\alpha'}{2}p'^{2}_{R}}  \left( 16\cdot 1+\mathcal{O}(q) \right)
+\sum_{m\in 2\boldsymbol{Z}+1}q^{\frac{\alpha'}{2}\left(\ell'^{2}_{L}+p'^{2}_{L} \right)} \bar{q}^{\frac{\alpha'}{2}p'^{2}_{R}}  \left( 16\cdot 14q^{\frac{1}{2}}+\mathcal{O}(q) \right) \right\rbrace,\\
&\bar{\eta}\eta^{-7}O^{(\alpha,\beta)}_{16}S_{16}=\bar{\eta}\eta^{-7}O^{(\alpha,\beta)}_{16}C_{16}\\
&=\sum_{n,w}\left\lbrace \sum_{m\in 2\boldsymbol{Z}}q^{\frac{\alpha'}{2}\left(\ell'^{2}_{L}+p'^{2}_{L} \right)}\bar{q}^{\frac{\alpha'}{2}p'^{2}_{R}}  \left(2^{8-1}\cdot 1+\mathcal{O}(q) \right)
+\sum_{m\in 2\boldsymbol{Z}+1}q^{\frac{\alpha'}{2}\left(\ell'^{2}_{L}+p'^{2}_{L} \right)} \bar{q}^{\frac{\alpha'}{2}p'^{2}_{R}}  \left( \mathcal{O}(q^{\frac{1}{2}}) \right) \right\rbrace,\\
&\bar{\eta}\eta^{-7}V^{(\alpha,\beta)}_{16}O_{16}
=q^{-1}\sum_{n,w}\left\lbrace \sum_{m\in 2\boldsymbol{Z}}q^{\frac{\alpha'}{2}\left(\ell'^{2}_{L}+p'^{2}_{L} \right)}\bar{q}^{\frac{\alpha'}{2}p'^{2}_{R}}  \left( 1\cdot 14 q^{\frac{1}{2}}+\mathcal{O}(q^{\frac{3}{2}}) \right)\right. \\ 
&\left.~~~~~~~~~~~~~~~~~~~~~~~~~~
+\sum_{m\in 2\boldsymbol{Z}+1}q^{\frac{\alpha'}{2}\left(\ell'^{2}_{L}+p'^{2}_{L} \right)} \bar{q}^{\frac{\alpha'}{2}p'^{2}_{R}} \left( 1+8q+\left(\frac{16\cdot 15}{2}+ \frac{14\cdot 13}{2}+1\right) q+\mathcal{O}(q^{\frac{3}{2}}) \right) \right\rbrace,\\
&\bar{\eta}\eta^{-7}V^{(\alpha,\beta)}_{16}V_{16}\\
&=q^{-\frac{1}{2}}\sum_{n,w}\left\lbrace \sum_{m\in 2\boldsymbol{Z}}q^{\frac{\alpha'}{2}\left(\ell'^{2}_{L}+p'^{2}_{L} \right)}\bar{q}^{\frac{\alpha'}{2}p'^{2}_{R}}  \left( 16\cdot 14 q^{\frac{1}{2}}+\mathcal{O}(q) \right)
+\sum_{m\in 2\boldsymbol{Z}+1}q^{\frac{\alpha'}{2}\left(\ell'^{2}_{L}+p'^{2}_{L} \right)} \bar{q}^{\frac{\alpha'}{2}p'^{2}_{R}} \left( 16\cdot 1 +\mathcal{O}(q) \right) \right\rbrace,\\
&\bar{\eta}\eta^{-7}V^{(\alpha,\beta)}_{16}S_{16}=\bar{\eta}\eta^{-7}V^{(\alpha,\beta)}_{16}C_{16}\\
&=\sum_{n,w}\left\lbrace \sum_{m\in 2\boldsymbol{Z}}q^{\frac{\alpha'}{2}\left(\ell'^{2}_{L}+p'^{2}_{L} \right)}\bar{q}^{\frac{\alpha'}{2}p'^{2}_{R}}  \left(\mathcal{O}(q^{\frac{1}{2}}) \right)
+\sum_{m\in 2\boldsymbol{Z}+1}q^{\frac{\alpha'}{2}\left(\ell'^{2}_{L}+p'^{2}_{L} \right)} \bar{q}^{\frac{\alpha'}{2}p'^{2}_{R}}  \left( 2^{8-1}\cdot 1+\mathcal{O}(q) \right) \right\rbrace,\\
&\bar{\eta}\eta^{-7}S^{(\alpha,\beta)}_{16}O_{16}=\bar{\eta}\eta^{-7}C^{(\alpha,\beta)}_{16}O_{16}
=q^{-\frac{1}{8}}\sum_{n,w}\sum_{m\in \boldsymbol{Z}+1/2}q^{\frac{\alpha'}{2}\left(\ell'^{2}_{L}+p'^{2}_{L} \right)}\bar{q}^{\frac{\alpha'}{2}p'^{2}_{R}}  \left( 1\cdot 2^{7-1}+\mathcal{O}(q) \right), \\ 
&\bar{\eta}\eta^{-7}S^{(\alpha,\beta)}_{16}V_{16}=\bar{\eta}\eta^{-7}C^{(\alpha,\beta)}_{16}V_{16}
=q^{\frac{3}{8}}\sum_{n,w} \sum_{m\in\boldsymbol{Z}+1/2}
q^{\frac{\alpha'}{2}\left(\ell'^{2}_{L}+p'^{2}_{L} \right)}\bar{q}^{\frac{\alpha'}{2}p'^{2}_{R}}  \left(\mathcal{O}(1) \right), \\ 
&\bar{\eta}\eta^{-7}S^{(\alpha,\beta)}_{16}S_{16}=\bar{\eta}\eta^{-7}C^{(\alpha,\beta)}_{16}S_{16}=\bar{\eta}\eta^{-7}S^{(\alpha,\beta)}_{16}C_{16}=\bar{\eta}\eta^{-7}C^{(\alpha,\beta)}_{16}C_{16}
=q^{\frac{7}{8}}\sum_{n,w} \sum_{m\in \boldsymbol{Z}+1/2}q^{\frac{\alpha'}{2}\left(\ell'^{2}_{L}+p'^{2}_{L} \right)}\bar{q}^{\frac{\alpha'}{2}p'^{2}_{R}}  \left(\mathcal{O}(1) \right).
\end{split}
\end{equation}
Note  that all states which come from $S^{(\alpha,\beta)}_{16}S_{16}$ ($=S^{(\alpha,\beta)}_{16}C_{16}=C^{(\alpha,\beta)}_{16}S_{16}=C^{(\alpha,\beta)}_{16}C_{16}$) or $S^{(\alpha,\beta)}_{16}V_{16}$ ($=C^{(\alpha,\beta)}_{16}V_{16}$) will never be massless.



\begin{thebibliography}{99}

\bibitem{String Theories in Ten-Dimensions Without Space-Time Supersymmetry} L.~J.~Dixon and J.~A.~Harvey,
	``String Theories in Ten-Dimensions Without Space-Time Supersymmetry,''
	Nucl.\ Phys.\ B {\bf 274}, 93 (1986).
	
\bibitem{An O(16) x O(16) Heterotic String} L.~Alvarez-Gaume, P.~H.~Ginsparg, G.~W.~Moore and C.~Vafa,
``An O(16) x O(16) Heterotic String,''
Phys.\ Lett.\ B {\bf 171}, 155 (1986).

\bibitem{Nonsupersymmetric orbifolds} A.~Font and A.~Hernandez,
``Nonsupersymmetric orbifolds,''
Nucl.\ Phys.\ B {\bf 634}, 51 (2002)
doi:10.1016/S0550-3213(02)00336-X
[hep-th/0202057].

\bibitem{Spontaneous Breaking of Supersymmetry Through Dimensional Reduction} J.~Scherk and J.~H.~Schwarz,
``Spontaneous Breaking of Supersymmetry Through Dimensional Reduction,''
Phys.\ Lett.\  {\bf 82B}, 60 (1979).

\bibitem{Spontaneous Supersymmetry Breaking in Supersymmetric String Theories} R.~Rohm,
``Spontaneous Supersymmetry Breaking in Supersymmetric String Theories,''
Nucl.\ Phys.\ B {\bf 237}, 553 (1984).

\bibitem{Supersymmetry Restoration in the Compactified O(16) x O(16)-prime Heterotic String Theory} H.~Itoyama and T.~R.~Taylor,
``Supersymmetry Restoration in the Compactified O(16) x O(16)-prime Heterotic String Theory,''
Phys.\ Lett.\ B {\bf 186}, 129 (1987).

\bibitem{Small Cosmological Constant in String Models} H.~Itoyama and T.~R.~Taylor,
``Small Cosmological Constant in String Models,''
FERMILAB-CONF-87-129-T.

\bibitem{Strong / weak coupling duality relations for nonsupersymmetric string theories} J.~D.~Blum and K.~R.~Dienes,
``Strong / weak coupling duality relations for nonsupersymmetric string theories,''
Nucl.\ Phys.\ B {\bf 516}, 83 (1998)
doi:10.1016/S0550-3213(97)00803-1 [hep-th/9707160].

\bibitem{Towards a nonsupersymmetric string phenomenology} S.~Abel, K.~R.~Dienes and E.~Mavroudi,
``Towards a nonsupersymmetric string phenomenology,''
Phys.\ Rev.\ D {\bf 91}, no. 12, 126014 (2015)
doi:10.1103/PhysRevD.91.126014 [arXiv:1502.03087 [hep-th]].

\bibitem{Interpolations from supersymmetric to nonsupersymmetric strings and their properties} B.~Aaronson, S.~Abel and E.~Mavroudi,
``Interpolations from supersymmetric to nonsupersymmetric strings and their properties,''
Phys.\ Rev.\ D {\bf 95}, no. 10, 106001 (2017)
doi:10.1103/PhysRevD.95.106001 [arXiv:1612.05742 [hep-th]].

\bibitem{Tension Between a Vanishing Cosmological Constant and Non-Supersymmetric Heterotic Orbifolds} S.~Groot Nibbelink, O.~Loukas, A.~Mütter, E.~Parr and P.~K.~S.~Vaudrevange,
``Tension Between a Vanishing Cosmological Constant and Non-Supersymmetric Heterotic Orbifolds,'' arXiv:1710.09237 [hep-th].

\bibitem{Exponential suppression of the cosmological constant in nonsupersymmetric string vacua at two loops and beyond} S.~Abel and R.~J.~Stewart,
``Exponential suppression of the cosmological constant in nonsupersymmetric string vacua at two loops and beyond,''
Phys.\ Rev.\ D {\bf 96}, no. 10, 106013 (2017)
doi:10.1103/PhysRevD.96.106013 [arXiv:1701.06629 [hep-th]]

\bibitem{Stability and vacuum energy in open string models with broken supersymmetry} S.~Abel, E.~Dudas, D.~Lewis and H.~Partouche,
``Stability and vacuum energy in open string models with broken supersymmetry,'' arXiv:1812.09714 [hep-th].

\bibitem{ATKIN-LEHNER SYMMETRY} G.~W.~Moore,
``Atkin-lehner Symmetry,''
Nucl.\ Phys.\ B {\bf 293}, 139 (1987)
Erratum: [Nucl.\ Phys.\ B {\bf 299}, 847 (1988)].

\bibitem{The Failure of Atkin-lehner Symmetry for Lattice Compactified Strings} 
J.~Balog and M.~P.~Tuite,
``The Failure of Atkin-lehner Symmetry for Lattice Compactified Strings,''
Nucl.\ Phys.\ B {\bf 319}, 387 (1989).

\bibitem{GENERALIZED ATKIN-LEHNER SYMMETRY} K.~R.~Dienes,
``Generalized Atkin-lehner Symmetry,''
Phys.\ Rev.\ D {\bf 42}, 2004 (1990).

\bibitem{Model Building on Asymmetric Z(3) Orbifolds: Nonsupersymmetric Models} T.~R.~Taylor,
``Model Building on Asymmetric Z(3) Orbifolds: Nonsupersymmetric Models,''
Nucl.\ Phys.\ B {\bf 303}, 543 (1988).

\bibitem{Non-supersymmetric Asymmetric Orbifolds with Vanishing Cosmological Constant} Yuji Satoh, Yuji Y.~Satoh, Y.~Sugawara and T.~Wada,
``Non-supersymmetric Asymmetric Orbifolds with Vanishing Cosmological Constant,''
JHEP {\bf 1602}, 184 (2016)
doi:10.1007/JHEP02(2016)184 [arXiv:1512.05155 [hep-th]].

\bibitem{More on Non-supersymmetric Asymmetric Orbifolds with Vanishing Cosmological Constant} Y.~Sugawara and T.~Wada,
``More on Non-supersymmetric Asymmetric Orbifolds with Vanishing Cosmological Constant,''
JHEP {\bf 1608}, 028 (2016)
doi:10.1007/JHEP08(2016)028 [arXiv:1605.07021 [hep-th]].

\bibitem{Vacuum energy cancellation in a nonsupersymmetric string} S.~Kachru, J.~Kumar and E.~Silverstein,
``Vacuum energy cancellation in a nonsupersymmetric string,''
Phys.\ Rev.\ D {\bf 59}, 106004 (1999)
doi:10.1103/PhysRevD.59.106004 [hep-th/9807076].

\bibitem{On vanishing two loop cosmological constants in nonsupersymmetric strings} S.~Kachru and E.~Silverstein,
``On vanishing two loop cosmological constants in nonsupersymmetric strings,''
JHEP {\bf 9901}, 004 (1999)
doi:10.1088/1126-6708/1999/01/004 [hep-th/9810129].

\bibitem{New Heterotic String Theories in Uncompactified Dimensions $<$ 10} K.~S.~Narain,
``New Heterotic String Theories in Uncompactified Dimensions < 10,''
Phys.\ Lett.\  {\bf 169B}, 41 (1986).

\bibitem{A Note on Toroidal Compactification of Heterotic String Theory} K.~S.~Narain, M.~H.~Sarmadi and E.~Witten,
``A Note on Toroidal Compactification of Heterotic String Theory,''
Nucl.\ Phys.\ B {\bf 279}, 369 (1987).

\bibitem{Toroidal Compactification of Nonsupersymmetric Heterotic Strings} P.~H.~Ginsparg and C.~Vafa,
``Toroidal Compactification of Nonsupersymmetric Heterotic Strings,''
Nucl.\ Phys.\ B {\bf 289}, 414 (1987).

\bibitem{Eternal Higgs inflation and the cosmological constant problem} Y.~Hamada, H.~Kawai and K.~y.~Oda,
``Eternal Higgs inflation and the cosmological constant problem,''
Phys.\ Rev.\ D {\bf 92}, 045009 (2015)
doi:10.1103/PhysRevD.92.045009 [arXiv:1501.04455 [hep-ph]].

\bibitem{A new twist on heterotic string compactifications} B.~Fraiman, M.~Graña and C.~A.~Núñez,
``A new twist on heterotic string compactifications,''
JHEP {\bf 1809}, 078 (2018)
doi:10.1007/JHEP09(2018)078 [arXiv:1805.11128 [hep-th]].

\bibitem{The Heterotic String} D.~J.~Gross, J.~A.~Harvey, E.~J.~Martinec and R.~Rohm,
``The Heterotic String,''
Phys.\ Rev.\ Lett.\  {\bf 54}, 502 (1985).

\bibitem{Casimir Effects in Superstring Theories} K.~Kikkawa and M.~Yamasaki,
``Casimir Effects in Superstring Theories,''
Phys.\ Lett.\  {\bf 149B}, 357 (1984).

\bibitem{Vacuum Energies of String Compactified on Torus} N.~Sakai and I.~Senda,
``Vacuum Energies of String Compactified on Torus,''
Prog.\ Theor.\ Phys.\  {\bf 75}, 692 (1986)
Erratum: [Prog.\ Theor.\ Phys.\  {\bf 77}, 773 (1987)].

\bibitem{Evidence for heterotic - type I string duality} J.~Polchinski and E.~Witten,
``Evidence for heterotic - type I string duality,''
Nucl.\ Phys.\ B {\bf 460}, 525 (1996)
doi:10.1016/0550-3213(95)00614-1 [hep-th/9510169]

\bibitem{Construction of Fermionic String Models in Four-Dimensions} H.~Kawai, D.~C.~Lewellen and S.~H.~H.~Tye,
``Construction of Fermionic String Models in Four-Dimensions,''
Nucl.\ Phys.\ B {\bf 288}, 1 (1987).

\bibitem{Four-Dimensional Superstrings} II.~Antoniadis, C.~P.~Bachas and C.~Kounnas,
``Four-Dimensional Superstrings,''
Nucl.\ Phys.\ B {\bf 289}, 87 (1987).

\bibitem{Classification of Closed Fermionic String Models} H.~Kawai, D.~C.~Lewellen and S.~H.~H.~Tye,
``Classification of Closed Fermionic String Models,''
Phys.\ Rev.\ D {\bf 34}, 3794 (1986).

\bibitem{Strings on Orbifolds} L.~J.~Dixon, J.~A.~Harvey, C.~Vafa and E.~Witten,
``Strings on Orbifolds,''
Nucl.\ Phys.\ B {\bf 261}, 678 (1985).

\bibitem{Strings on Orbifolds. 2.} L.~J.~Dixon, J.~A.~Harvey, C.~Vafa and E.~Witten,
``Strings on Orbifolds. 2.,''
Nucl.\ Phys.\ B {\bf 274}, 285 (1986).

\bibitem{Itoyama:1987qz} 
H.~Itoyama and P.~Moxhay,
``Multiparticle Superstring Tree Amplitudes,''
Nucl.\ Phys.\ B {\bf 293}, 685 (1987).

\bibitem{Itoyama:2015fht} 
H.~Itoyama and K.~Yano,
``Genus one super-Green function revisited and superstring amplitudes with non-maximal supersymmetry,''
PTEP {\bf 2016}, no. 3, 033B05 (2016)
doi:10.1093/ptep/ptw019 [arXiv:1512.07705 [hep-th]].

\bibitem{Arakane:1996rh} 
Y.~Arakane, H.~Itoyama, H.~Kunitomo and A.~Tokura,
``Infinity cancellation, type I-prime compactification and string S matrix functional,''
Nucl.\ Phys.\ B {\bf 486}, 149 (1997)
doi:10.1016/S0550-3213(96)00655-4 [hep-th/9609151].

\bibitem{Kostelecky:1986xg} 
V.~A.~Kostelecky, O.~Lechtenfeld, W.~Lerche, S.~Samuel and S.~Watamura,
``Conformal Techniques, Bosonization and Tree Level String Amplitudes,''
Nucl.\ Phys.\ B {\bf 288}, 173 (1987).

\bibitem{Kounnas:2016gmz} 
C.~Kounnas and H.~Partouche,
``Super no-scale models in string theory,''
Nucl.\ Phys.\ B {\bf 913}, 593 (2016)
doi:10.1016/j.nuclphysb.2016.10.001
[arXiv:1607.01767 [hep-th]].

\bibitem{Kounnas:2017mad} 
C.~Kounnas and H.~Partouche,
``$\mathcal N=2 \to 0$ super no-scale models and moduli quantum stability,''
Nucl.\ Phys.\ B {\bf 919}, 41 (2017)
doi:10.1016/j.nuclphysb.2017.03.011
[arXiv:1701.00545 [hep-th]].

\bibitem{Florakis:2016ani} 
I.~Florakis and J.~Rizos,
``Chiral Heterotic Strings with Positive Cosmological Constant,''
Nucl.\ Phys.\ B {\bf 913}, 495 (2016)
doi:10.1016/j.nuclphysb.2016.09.018
[arXiv:1608.04582 [hep-th]].
\end{thebibliography}
\end{document}